\documentclass[floats,floatfix,showpacs,preprint,prd,superscriptaddress,onecolumn,nofootinbib]{revtex4}
\usepackage{graphicx, epsfig, amssymb} 
\usepackage{amsmath, amsfonts}
\usepackage{bm} 
\usepackage{color}
\usepackage[usenames]{xcolor}

\newcommand{\be}{\begin{equation}}
\newcommand{\ee}{\end{equation}}                  
\newcommand{\bea}{\begin{eqnarray}}
\newcommand{\eea}{\end{eqnarray}}

\begin{document}

 
\title{Boson stars in biscalar extensions of Horndeski gravity}

\author{Yves Brihaye}
\email{yves.brihaye@umons.ac.be}
\affiliation{Theoretical and Mathematical Physics Department, University of Mons 20,\\Place du Parc - 7000 Mons, Belgium}
\author{Adolfo Cisterna}
\email{adolfo.cisterna@uach.cl}
\affiliation{Instituto de Ciencias F\'isicas y Matem\'aticas,\\ Universidad Austral de Chile,\\ Valdivia, Chile}
\affiliation{Universidad Central de Chile, Santiago, Chile.}
\author{Cristi\'an Erices}
\email{erices@cecs.cl}
\affiliation{Departamento de F\'isica, Universidad de Concepci\'on,\\
Casilla, 160-C, Concepci\'on, Chile}
\affiliation{Centro de Estudios Cient\'{\i}ficos (CECs), Casilla 1469, Valdivia, Chile.}

\begin{abstract}
This paper is concerned with the construction and analysis of boson stars in the context of nonminimal derivative coupling theories. In particular we embed our model in the biscalar extension of Horndeski gravity, considering a scalar field theory displaying a nonminimally coupled kinetic term given by the Einstein tensor. We focus on the case where the potential is given by a mass term only, and when a six order self-interaction is included. In the latter case we consider specific couplings in the self-interacting terms in such a way that our self-interaction is given by a positive definite potential presenting two degenerate local vacua. We show how solutions can be obtained and we compare its principal properties with standard configurations constructed with the usual minimally coupled kinetic term. 
\end{abstract}

\maketitle

\section{Introduction}

Recently, scalar-tensor theories of gravity (STT) have attracted huge attention. Among the simplest candidates to represent a viable modified gravity theory, STT aim to describe the gravitational interaction considering extra degrees of freedom, in particular scalar fields. The prototype example of STT is the pioneer work of Brans and Dicke \cite{BD}, where gravity is described by a massless spin-two particle and a single real scalar field. All matter fields are coupled only to the metric tensor in order to avoid possible violations of the weak equivalence principle. This is a common assumption in STT, that promotes them as effective field theories of some more fundamental theory, as should be a quantum description of the gravitational interaction, in which such couplings should not be neglected. \\
The most general STT with second order equations of motion, constructed with a single real scalar field, a single metric tensor and with a Levi-Civita connection, was given by Horndeski in the early seventies \cite{Horndeski}. However, overlooked for a while, Horndeski theory attracted again attention after the appearance of Galileon theories \cite{Nicolis}, a scalar field theory originally motivated by the decoupling limit of the Dvali-Gabadadze-Porrati (DGP) model \cite{DGP}. The DGP model is a novel higher dimensional modification of gravity based on the existence of a five-dimensional Minkowski spacetime in which a 3-brane surface containing all matter fields is embedded. Contrary to other higher dimensional models, in the DGP scenario the extra dimensions have an infinite size.

Along with the usual Einstein-Hilbert action in five dimensions, the DGP model also possesses a curvature term on the brane, given by the four-dimensional Einstein-Hilbert action, induced by the attached matter fields. From a four-dimensional point of view, gravity is mediated by a massive spin-two particle and a scalar degree of freedom. The standard gravitational potential is recovered at small distance scales, while a fully 5D potential dominates when the scales are larger than a specific crossover limit, representing an explicit infrared (IR) modification of gravity.
This model was extensively analyzed due to its interesting cosmological solutions \cite{Deffayet:2000uy, Lue:2005ya, Deffayet:2001pu}. In particular, it was shown that one solution branch possesses a self-accelerating behavior without any cosmological constant term.\\
The decoupling limit of the DGP model contains a very appealing effective scalar field theory. The resulting Lagrangian possesses higher order terms that combine in such a way that the resulting equations of motion are of second order. Moreover, the theory also contains a proper screening mechanism, the Vainshtein mechanism \cite{Babichev:2013usa}  and is invariant under Galilean transformations \footnote{Galilean transformations are given by $\phi\rightarrow\phi+\phi_0+b_\mu x^\mu$, with $\phi_0$ and $b_\mu$ constants.}. Soon after these developments, the DGP decoupling limit was generalized to the most general set of Lagrangians sharing the same properties, namely, Galileon theories \cite{Nicolis}. It was shown in \cite{galileon1}, that following the standard approach of covariantization, the resulting theory has equations of motion of third order. In the same work, the authors develop the procedure to construct Galileon theories on curved spacetimes conserving the second order character of the theory. To do so, several nonminimal couplings between the scalar field and curvature terms must be added as counter terms that exactly cancel all higher derivative contributions. The D-dimensional version of the theory is given in \cite{Deffayet:2009mn}. The covariant version of Galileon gravity was demonstrated to be equivalent to the original Horndeski theory \cite{Kobayashi:2011nu}. Its Lagrangian can be cast in the very explicit form
\begin{equation}
S=\sum_{i=2}^{5}\int d^{4}x\sqrt{-g}{\cal L}_i\ , 
\label{gal}
\end{equation}
where
\begin{eqnarray*}
{\cal L}_2&=&G_2\ ,\\
{\cal L}_3&=&-G_{3}\square\phi\ ,\\
{\cal L}_4&=&G_{4}R+G_{4{X}}\left[(\square\phi)^2-(\nabla_{\mu}\nabla_{\nu}\phi)^2\right]\ ,\\
{\cal L}_5&=&G_{5}G_{\mu\nu}\nabla^{\mu}\nabla^{\nu}\phi-\frac{G_{5{X}}}{6}\left[(\square\phi)^3+2(\nabla_{\mu}\nabla_{\nu}\phi)^3-3(\nabla_{\mu}\nabla_{\nu}\phi)^2\square\phi\right]\ .
\end{eqnarray*}
Here, $G_i$ are arbitrary functions of the scalar field and of its standard canonical kinetic term $X\equiv -\nabla_{\mu}\phi\nabla^{\mu}\phi$, while $G_{iX}$ stands for their derivatives with respect to $X$.
\\
As we can see, this Lagrangian possesses a huge freedom encoded in the arbitrary functions $G_i$. One very nice phenomenological requirement we can impose is to consider which is the most general sector of Horndeski gravity that allows for self-tuning cosmological solutions. This means, which sector of Horndeski theory should admit Minkowski spacetime as a solution in the presence of an arbitrary cosmological constant (any value of the vacuum energy). 
This was carried out in \cite{fabfour}, where the authors, permitting the scalar field to evolve in time, circumvent Weinberg's theorem \cite{{Weinberg:1988cp}} allowing the scalar field equation to force the vacuum curvature to vanish.\\
In the context of inflationary cosmology and perturbation theory, the nonminimal kinetic coupling sector ${\cal L}_5$ has received lot of attention, see e.g \cite{Germani:2010gm, Amendola:1993uh, Sushkov:2009hk, Myrzakulov:2015ysa, Gumjudpai:2015vio, namur} and references therein.\\
It was demonstrated that in the case where $G_{5}=G_{5}(\phi)$ ($G_{5X}=0$) accelerating behaviors are obtained without including any potential term. In perturbation theory a lot of work was done in order to find potentially observable deviations from GR in large-scale structures and the conditions on the parameter space that avoid too large gravitational instabilities \cite{pert}. For works considering the whole Horndeski Lagrangian see \cite{Amendola:2012ky, Motta:2013cwa}.\\
Including the canonical kinetic term, this model can be represented by the following action:
\begin{equation}
S=\int d^{4}x\sqrt{-g}\left[\left(\frac{R-2\Lambda}{16\pi G}\right)-\frac{1}{2}\left(\alpha g^{\mu\nu}-\eta G^{\mu\nu} \right)\nabla_{\mu}\phi\nabla_{\nu}\phi \right], \label{model}
\end{equation}
where $R$ is the Ricci scalar, $G$ denotes Newton's constant and $\alpha$ and $\eta$ are two parameters controlling the strength of the minimal and nonminimal kinetic couplings, respectively. Along this paper, we use natural units where $\hbar=c=1$.
\\
Black hole solutions and neutron star configurations have also been investigated in this model. Constrained at the beginning by the existence of a no-hair theorem \cite{Hui:2012qt}, Horndeski black hole solutions do exist in the nonminimal kinetic sector when static asymptotically AdS geometries are considered \cite{rinaldi, Kolyvaris:2011fk, charmousis1, Kobayashi:2014eva, adolfo2, Minam1, adolfo3, MK, minas}. The stability of solutions with a linear time-dependent scalar field has been recently studied in \cite{Ogawa}. Slowly rotating configurations have also received attention \cite{Maselli:2015yva, Cisterna:2015uya}. Moreover, recently the study of how these configurations form through gravitational collapse has been addressed \cite{Koutsoumbas:2015ekk}.\\ 
The construction of neutron stars has been tackled first in \cite{Cisterna:2015yla}. There, static neutron stars and white dwarfs are shown to be supported for the particular case in which the scalar field is kinetically coupled only through the Einstein tensor ($\alpha=0$), imposing in a very natural way, astrophysical constraints on the only free parameter that these solutions exhibit. Progress on slowly rotating neutron stars are provided in \cite{Silva:2016smx} considering polytropic equations of state and in \cite{Cisterna:2016vdx} considering realistic equation of state tables describing nuclear matter. In this latter work, the authors have constructed neutron stars with the maximum observed mass to date for these kind of objects, namely the mass of the pulsar PSR J0348+0432 \cite{Demorest:2010bx} ($2.01 \pm 0.04 M_\odot$ and with an orbital period of $2$ hours and $27$ minutes). 
Based on the constraints imposed in \cite{Cisterna:2015yla}  they show for a set of tabulated equations of state which of them can or cannot satisfy the desired mass bound and how the sign of the nonminimal coupling parameter affects the masses of these configurations. Recently in \cite{Maselli:2016gxk} the authors have followed the same procedure to understand if the construction of these neutron stars can be made in other subsectors of the Lagrangian (\ref{gal}). In particular, it has been shown that the Paul sector of the theory \cite{fabfour} does not support these kindsnontopological of compact objects.


In the present paper we discuss the construction of nontopological soliton solutions for the model (\ref{model}), in particular the gravitating solitons known as boson stars (BSs), solutions that in many aspects behave as neutron star configurations.
As far as our we know, this problem has not been tackled previously in the literature.\\
Solitons are solutions of the nonlinear equations of motion of field theories and they represent localized particle-like objects with finite energy. These solutions can be interpreted as particle of the theory under consideration, but remarkably different from the standard quantum field theory particles. Solitons possess a nontrivial topological structure which is precisely the responsible of their stability \cite{Manton:2004tk}.\\
Contrary to what happens for topological solitons, the stability of nontopological solitons is due to the existence of a globally conserved Noether charge. This Noether charge is the result of a global internal continuous symmetry of the system. This charge can be promoted to a local Noether charge when considering gauge field theories.\\
BSs originally constructed in \cite{Kaup:1968zz} are compact stationary solutions of the Einstein-Klein-Gordon equations with a complex scalar field configuration. These solutions, which have shown the possibility to be stable \cite{Kusmartsev:1990cr, Kleihaus:2011sx}, represent a balance between the attractive nature of gravity and the dispersive behavior of scalar fields and can be thought as a collection of stable fundamental scalar fields bounded by gravity, where the Noether charge represents the total number of bosonic particles. Contrary to what occurs with nongravitating solitons in this context, namely, Q-balls \cite{Lee:1991ax}, BSs do not need a self-interaction. Rotating and nonrotating Q-balls have been shown to exist with polynomial self-interaction containing up to six-th order terms \cite{Volkov:2002aj, Kleihaus:2005me}. Also studied is the case where this kind of localized object exists for self-interacting potential motivated by supersymmetric extensions of the standard model \cite{Kusenko:1997zq, Kusenko:1997ad, Hartmann:2012gw}. 

BSs can be constructed even with scalar fields possessing only a mass term \cite{Friedberg:1986tq, Jetzer:1991jr, Liddle:1993ha}. Due to bounds on the maximal mass of these configurations, they are usually referred as mini-BSs. For instance mini-BSs have a maximal mass of the order $M_{max}=0.6\frac{M^{2}_{Planck}}{m^2}$ where the Planck mass is $M_{Planck}=\sqrt{\frac{hc}{2\pi G}}$ and $m$ represents the mass of the bosonic particles. On this setup BSs have been proposed as possible constituents of dark matter halos, that could give an explanation of the anomaly rotation curve of some galaxies \cite{Kusenko:1997si, Eby:2015hsq}.\\ 
In \cite{shapiro} it was found that BSs with masses of the order of astrophysical objects can be obtained including self-interactions without collapsing into black holes. For the particular case of quartic self-interactions ($\lambda |\phi|^{4}$) the masses of the BSs can reach astrophysical orders, 
$M_{max}=(0.1 GeV^{2})M_{\odot}\frac{\sqrt{\lambda}}{m^2}$ depending on the specific value of the coupling parameter $\lambda$. This mass is of the same order as the Chandrasekhar mass for fermions \cite{shapiro}.\\ 
In this scenario BSs have been proposed as possible candidates to represent supermassive objects at the center of galaxies, and it is expected due to their dynamics to be able to be detected by astronomical observations \cite{Vincent:2015xta, Torres:2000dw, Macedo:2013jja, Meliani:2015zta, Cunha:2015yba}. Indeed astronomical observations of the center of nearby galaxies suggest the presence of a single large mass at their galactic center \cite{Ghez}. Although it has been widely accepted that this object is a supermassive black hole, the existence of an event horizon has only been inferred indirectly and not conclusively proved. Consequently, many candidates have emerged as alternative compact objects such as soliton stars \cite{Friedberg:1986tq} and neutrino balls \cite{TV}. In this context BSs have strongly drawn attention \cite{RB}. BSs have been shown to exist in several models including rotation, cosmological constant, in some modified gravity theories, higher dimensional scenarios and also containing fermionic matter contributions \cite{hartmann_riedel, Herdeiro:2014goa, Herdeiro:2015tia, Astefanesei:2003qy, Brihaye:2009yr} \footnote{Two self-contained reviews are \cite{Liebling:2012fv, Schunck:2003kk}.}.

It has been shown that observational properties of BSs are quite similar to its counterpart in black holes. Even more, the detection of a shadow and its photon ring in the strong field region is not definitive proof of the existence of an event horizon \cite{Vincent:2015xta}. Therefore, to unambiguously differentiate a black hole from a BS the detection is required of relativistic velocities in orbits of the order of a few Schwarzschild radii. It means that astronomical observations require resolutions currently unavailable, and we expect that, in the near future, the development of the Very Long Baseline Interferometry or the Event Horizon Telescope could enlighten us about the definitive answer \cite{Lu:2014zja, baseline}.

This paper is organized as follows: the next section is devoted to present our model and the general setting in which we study BS configurations. In Section \ref{sec3} we construct BSs considering no self-interaction. This means considering only a mass term, and we compare them with the results obtained for the standard case of minimally coupled scalar field theories. Section \ref{sec4} considers the inclusion of self-interaction, in particular the sixth order potential with nontrivial vacuum manifold. Finally we conclude in Sec. \ref{sec5}.

\section{General setting}
\subsection{The model}
In the following we extend (\ref{model}) to contain a complex scalar field. The action then reads
\begin{equation}
S=\int d^4x\sqrt{-g}\left(\frac{R}{16\pi G_N}\right) - \int{d^4x\sqrt{-g}[(\alpha g^{\mu\nu}-\eta G^{\mu\nu})\nabla_{\mu}\Phi\nabla_{\nu}\Phi^*+U(\vert\Phi\vert)}]  \label{lag}
\end{equation} 
where $\Phi$ denotes a complex scalar field. We work in the $(-+++)$ signature. As we mention above $\alpha$ and $\eta$ are the dimensionful parameters controlling the standard and nonminimal couplings. 
The potential $U(\vert\Phi\vert)$ contains the mass term $m$ and, eventually, a self-interaction to be specified below.\\
To embed this model in the context of the STT we are considering here, it is necessary to go beyond the original Horndeski theory and to consider its biscalar extension. Indeed, as we know, a system composed by a complex scalar field can be treated as a system composed by two real scalar fields. Extensions of Galileon theory or Horndeski gravity for the case in which two scalar fields degrees of freedom are considered have already been constructed in \cite{Padilla:2012dx, Ohashi:2015fma, Kobayashi:2013ina, Padilla:2010ir, Padilla:2010de}. We observe that in the biscalar extension also appears the nonminimal kinetic sector described above in (\ref{gal}) and that our model can be supported by that kind of Lagrangians. Construction of relativistic stars on these kinds of models have been considered in \cite{Osilva}. We point out that these kinds of theories have been recently considered in cosmology \cite{Cognola, Saridakis:2016ahq, Saridakis:2016mjd} where the authors have studied theories beyond Horndeski (higher order terms) imposing conformal invariance, thus arriving to a healthy ghost-free biscalar tensor theory. 

\subsection{The Ansatz}
Due to the complexity of the equations, we limit ourselves to stationary nonspinning solutions.
For this purpose we use a spherically symmetric ansatz for the metric and specify the radial variable through the isotropic coordinates
\begin{equation}
\label{metric}
 ds^2=-F(r) dt^2 + \frac{G(r)}{F(r)}\left[dr^2 + r^2 d\theta^2 + r^2 \sin^2\theta d\varphi^2\right]  \ .
\end{equation}
The scalar field is given by
\begin{equation}
\label{ansatz1}
\Phi= \Phi_0 \phi(r) e^{i \tilde \omega t}  \
\end{equation}
where the constant $\Phi_0$ supports the dimension of the scalar field and the frequency $\tilde \omega$ encodes the harmonic dependence of the solution. The harmonic ansatz is used in order to circumvent Derrick's theorem \cite{Derrick}, which states that time-independent localized solutions of nonlinear wave equations in spacetime with three or more space dimensions are unstable. For this precise form of the scalar field we obtain that the contribution of the scalar field in the equations of motion remains static, even if the scalar field degree of freedom is no longer static, not sharing in this way the same symmetries than the spacetime. 
The coupled system of nonlinear equations then reads
\begin{eqnarray} 
\label{equation_form}
& A_{11} F'' + A_{12} G'' + A_{13} \phi'' &= K_1(F,F',G,G',\phi, \phi',\tilde \omega) \nonumber \\
& A_{21} F'' + A_{22} G'' + A_{23} \phi'' &= K_2(F,F',G,G',\phi, \phi',\tilde \omega) \nonumber \\
& A_{31} F'' + A_{32} G'' + A_{33} \phi'' &= K_3(F,F',G,G',\phi, \phi',\tilde \omega)  \label{system}
\end{eqnarray} 
where the prime denotes the derivative with respect to $r$. Here, $K_a$ are polynomials given in term of the metric functions, the scalar field and their first derivatives respectively.
The coefficients $A_{ab}$ depends on the fields in the same way as the polynomials $K_a$. They are given in the Appendix.

In the case of a minimal coupling, i.e. for $\eta = 0$, the matrix $A$ is diagonal
and positive definite. Nevertheless, for $\eta \neq 0$, this matrix becomes 
nondiagonal and the determinant $|{\rm det} A(r)|$ plays a fundamental role in the existence of solutions. 
When this determinant presents zeros  the corresponding system is  singular
and no regular solution can be found. We see that this affects significantly the pattern of solutions.
 
\subsection{Boundary conditions}
For the construction of BSs, the system has to be solved with the following boundary conditions:
\be
\label{boundary_conditions}
\begin{split}
&F(0)=1\ ,\ G(0)=1\ ,\ \phi(0)=\phi_0\ ,\ F'(0)=0\ ,\ G'(0)=0\ ,\ \phi'(0)= 0\ ,\\&F(\infty)=1\ ,\ G(\infty)=1\ ,\ \phi(\infty)=0\ .
\end{split}     
\ee
Here $\phi_0$ represents the central value of the scalar field. On the one hand, the conditions at $r=0$ are necessary for soliton solutions to be regular at the origin. On the other hand the conditions at $r = \infty$ ensure localized and asymptotically flat solutions. To find solutions respecting these 
conditions on $r=0$ and $r=\infty$, 
the eigenvalue $\tilde \omega$  has to be fine-tuned for a given central value, $\phi_0$, of the scalar function $\phi$.
This  leads, in general, to a relation of the form $\omega(\phi_0)$. 
In principle, the equations can be solved by a shooting technique; we used instead the routine Colsys \cite{colsys}
 based on the Newton-Raphson algorithm.

\subsection{Rescaling}\label{rescaling}
For the numerical study of our system (\ref{equation_form}), it is convenient to perform suitable rescalings of the parameters leading to dimensionless quantities. For this purpose, we define the dimensionless variable $x$ and parameters $\kappa,\xi, \omega$ by means of
\be
    x = m r \ \ , \ \     \kappa = 8 \pi G_N \Phi_0^2 \ \ , \ \ \xi = \frac{\eta}{m^2} \ \ , \ \   \omega=\frac{\tilde \omega}{m}\ .  
\ee
where $m$ denotes the mass of the scalar field.
One reason for including the parameter $\alpha$ is to allow -if they would exist-  exotic 
 solutions corresponding to $\alpha=0$ and $\eta = 1$.
Since we failed to construct such solutions in the model under consideration,
we set, without losing generality, $\alpha = 1$  throughout the paper.
\subsection{Physical Quantities}
The solutions can be characterized by several quantities. 
The global symmetry of the action under
phase change of the scalar field leads to a conserved current $j^{\mu}$ and conserved charge $Q$:
\begin{equation}
j^{\mu} = - i (\Phi^* \partial^{\mu} \Phi - (\partial^{\mu} \Phi^*) \Phi) \ \ , \ \ Q_{phys} =  - \int j^0 \sqrt{-g} d\Sigma^3\ ,
\end{equation}
where $d\Sigma^3$ stands for the line element of the three-dimensional spatial hypersurface. With the ansatz  and rescaling used above,   
the conserved charge is  computed as follows
\begin{equation}
 Q_{phys}= 8\pi \frac{\Phi_0^2}{m^2} \int\limits_{0}^{\infty} \frac{\sqrt{G^3}}{F^2} x^2 \omega \phi^2 dx \ \ , \ \ 
  \frac{\Phi_0^2}{m^2} = \kappa \frac{M_{Pl}^2}{m^2} \ ,
\end{equation}
where $8 \pi G_N \equiv M_{Pl}^{-2}$ is the Planck mass;
the quantity $Q$ is interpreted as the number of bosonic particles. The solution is also characterized by 
the mass $M$, it can be read out of the asymptotic decay of the metric function $F$
\be
\label{mass}
        F(r) = 1 - \frac{2G_N M_{phys}}{r} + \mathcal{O}\left(\frac{1}{r^2}\right) \  = \  1 - \frac{2m M_{phys}}{8 \pi M_{Pl}^2}\frac{1}{x} + \mathcal{O}\left(\frac{1}{x^2}\right).
\ee 
The quantities $Q$ and $M$ reported  on the figures are related to the physical quantities according to
\be
     Q_{phys}=\kappa \frac {M_{Pl}^2}{m^2} Q \ \ \ , \ \ \ M_{phys} =   \kappa \frac {M_{Pl}^2}{m} M\ .  
\ee
The BS can also be characterized by a radius. There are
many ways to define such a parameter since the scalar field
does not strictly vanish, along many authors (see namely \cite{Pugliese:2013gsa})
we define the dimensionless radius $R$ of the BS as
\be
                \frac{R}{m} = \frac{1}{Q_{phys}} \int  r \ j^0 \sqrt{-g} d\Sigma^3 \ \ .
\ee
We find $R$ of order one; as a consequence,
a mass $m$ for the boson field of order one MeV  would corresponds to $R_{phys}$ of order 200 Fermi.
The ratio $M/mQ$ provides some information about the stability of the soliton. The condition $M < m Q$ is necessary for the soliton to be stable. Indeed if $M > m Q$ the mass of the full soliton exceeds the mass of $Q$ scalar field quanta and no binding energy is left to stabilize the lump.  In the discussion  of the solutions we refer to this argument only; the full study of the stability is out of the scope of this paper.  

 \subsection{The potentials}
 BS solutions minimally coupled to Einstein gravity with no self-interaction (mass term only) have been studied in great detail in \cite{Pugliese:2013gsa}.
 As we pointed out, the nongravitating counterpart of BSs, Q-balls, do not exist. Indeed, to obtain the later configurations is necessary to consider self-interaction with at least sixth order powers of the scalar field (see \cite{Volkov:2002aj}). Motivated by this, we also investigate BSs in the context of this kind of potentials for our nonminimally coupled model. The potential reads
\be
        V = \lambda_3 |\Phi|^6 - \lambda_2 |\Phi|^4 + \lambda_1 |\Phi|^2 \  \ , \ \ \lambda_1 \equiv m^2.
\ee
Because of the numerous parameters, we put the emphasis on the following two cases:
\begin{itemize}
\item  $\lambda_2=0$, $\lambda_3=0$ which corresponds to a mass term only. We examine  the influence of the nonminimal coupling on the spectrum of the solutions.
\item  $\lambda_2/\lambda_1= 2 \lambda_3/\lambda_1 = 2$. This corresponds to a positive
definite potential presenting two degenerate local minima at $\phi = 0$ and $\phi = 1$. We denote it as $V_6$.
\end{itemize}

\section{Boson stars with the  mass potential}\label{sec3}
As pointed out already, the occurrence of nodes of the quantity $| {\rm  det} A(r,\xi)|$ plays a role in the 
construction of the solutions. 
For all parameters that we have explored, the minimum of this determinant is always 
located at the origin (i.e.  $x=0$). Therefore  we find it convenient to define 
\be
          \Delta(\xi) = \frac {{\rm  det} A(0,\xi)}{{\rm  det} A(0,0)}     \label{control}
\ee
as a control parameter. The set of numerical routines employed
lead to reliable solutions  as long as $\Delta > 10^{-6}$.\\

\subsection{Mini-boson stars with $\xi=0$}
\begin{figure}[ht!]
\begin{center}
{\label{fig_1_l}\includegraphics[width=8.cm]{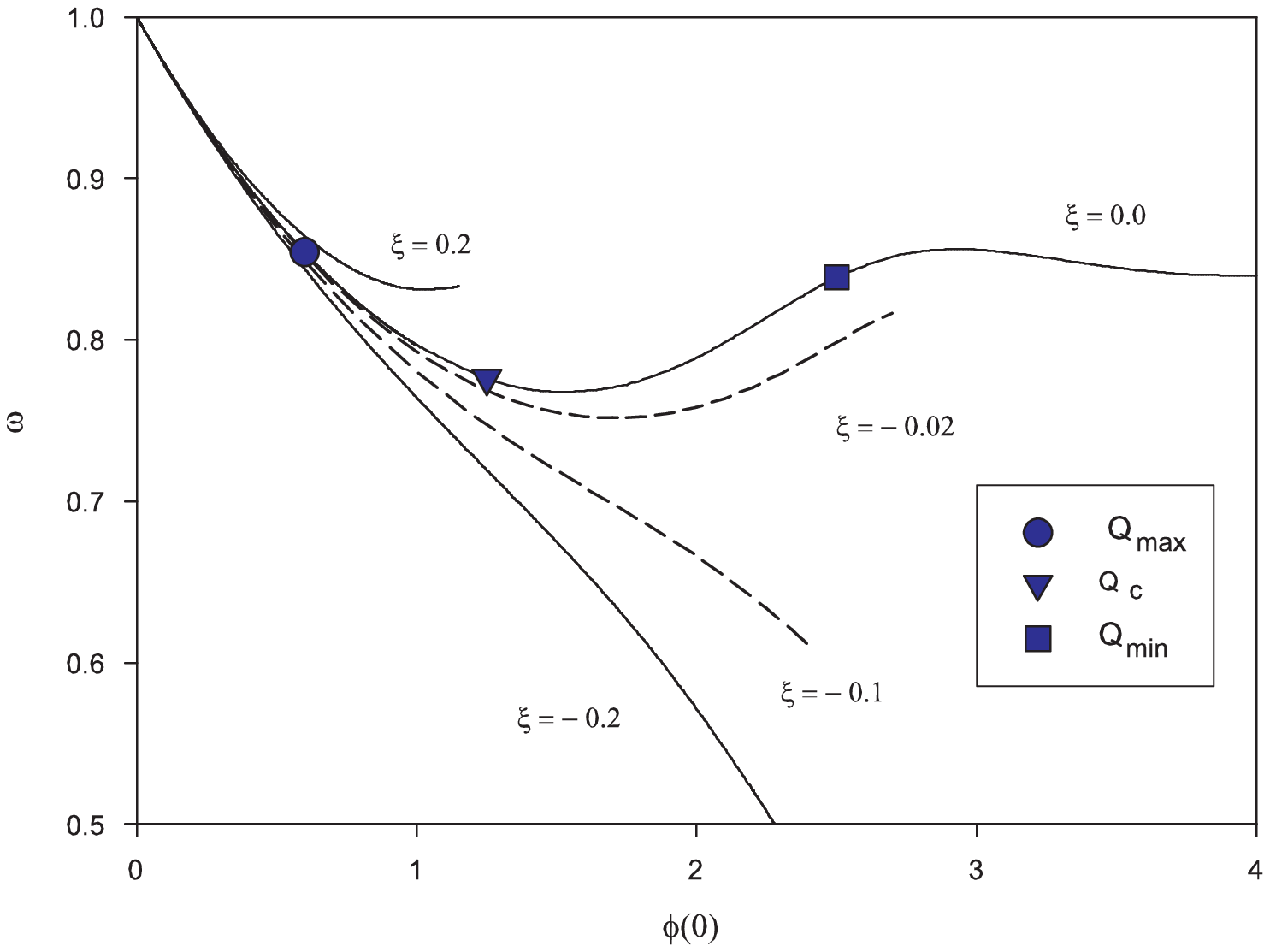}}
{\label{fig_1_r}\includegraphics[width=8.cm]{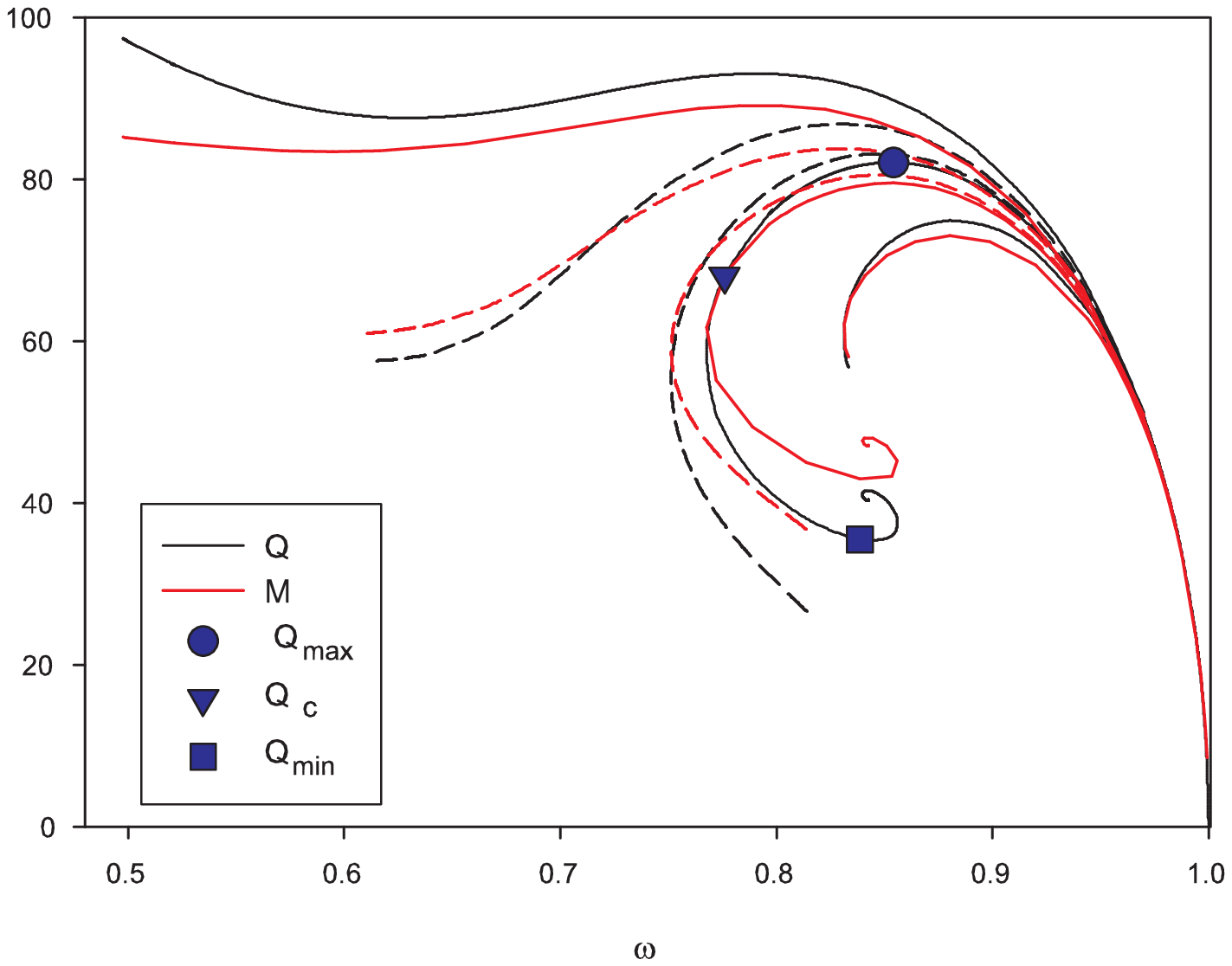}}
\end{center}
\caption{Left: The frequency $\omega$ as function of $\phi_0$ for BSs without self-interacting scalar fields and
for different values of  $\xi$. Right:
The mass and charge  as functions of $\omega$ for the same values of $\xi$. The three symbols (bullet, triangle and square)
show three critical values of $Q$ on the $\xi=0$ line. 
\label{fig_1}
}
\end{figure}

\begin{figure}[ht!]
\begin{center}
{\label{fig_1_l}\includegraphics[width=8.cm]{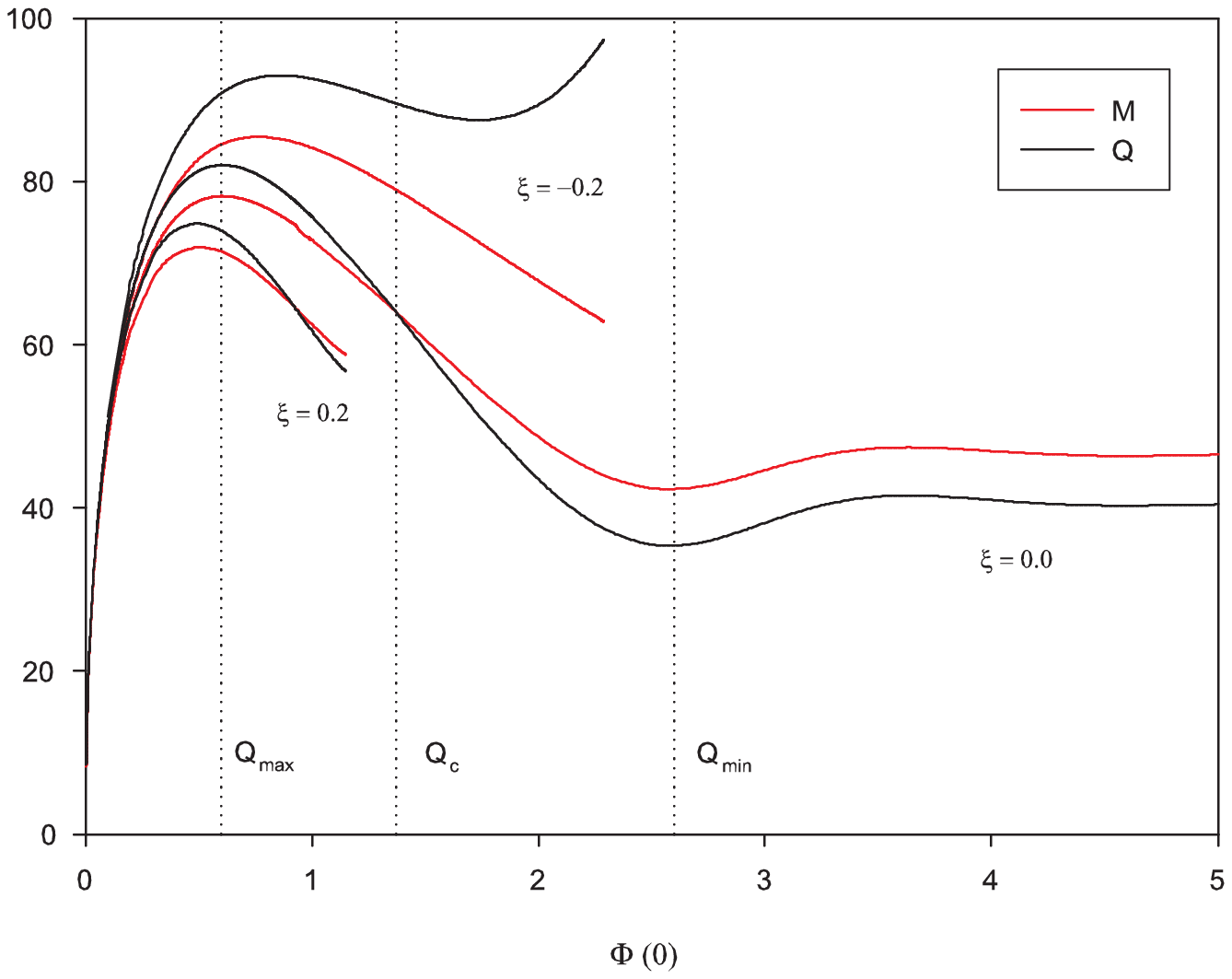}}
{\label{fig_1_r}\includegraphics[width=8.cm]{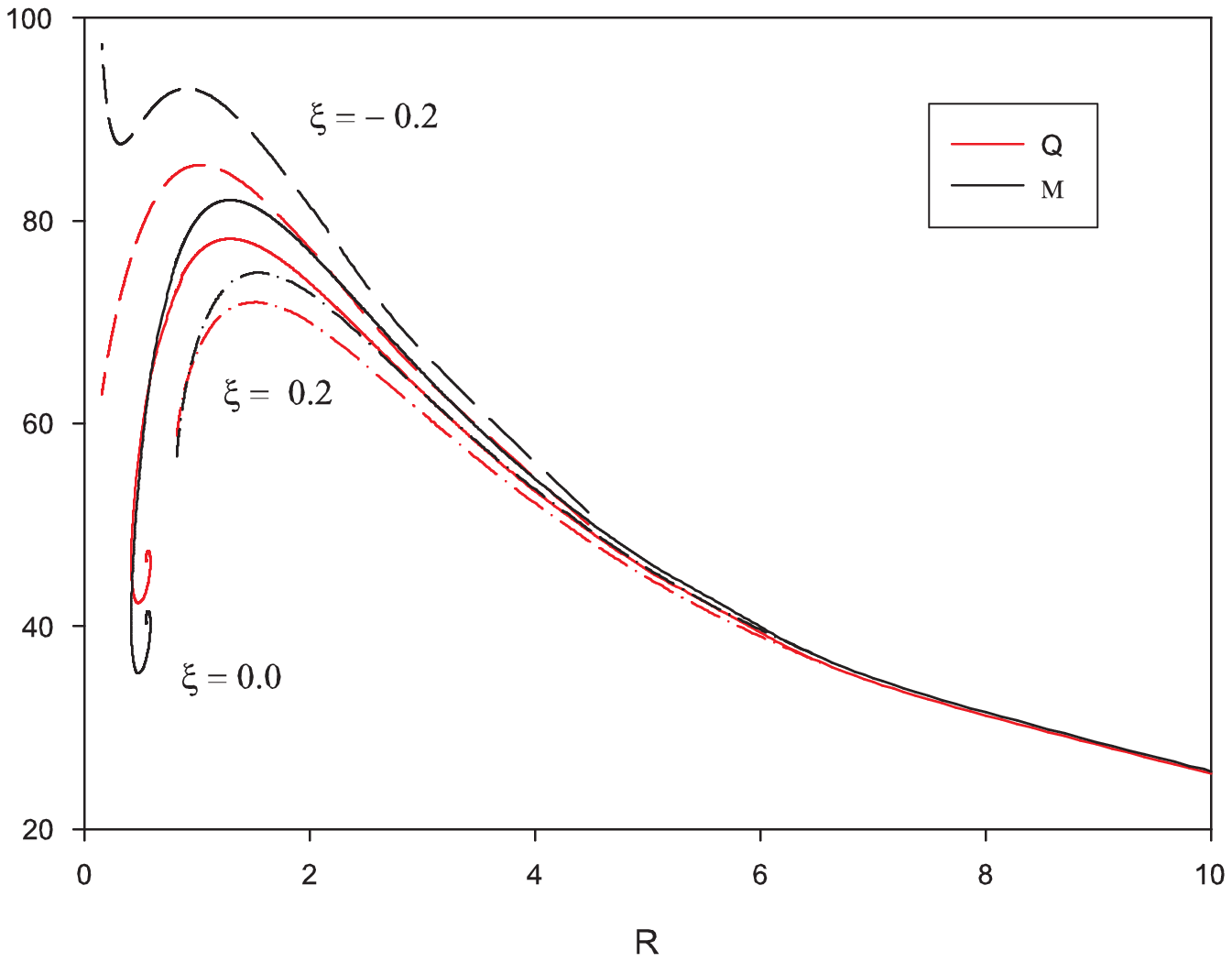}}
\end{center}
\caption{Left: The mass and charge as functions of $\phi_0$ for $\xi=0$ and $\xi = \pm 0.2$ . Right:
The mass and charge  as functions of the radius $R$ for the same values of $\xi$. 
\label{fig_1_new}
}
\end{figure}

\begin{figure}[h]
\begin{center}
{\label{fig_1_bis_l}\includegraphics[width=8cm]{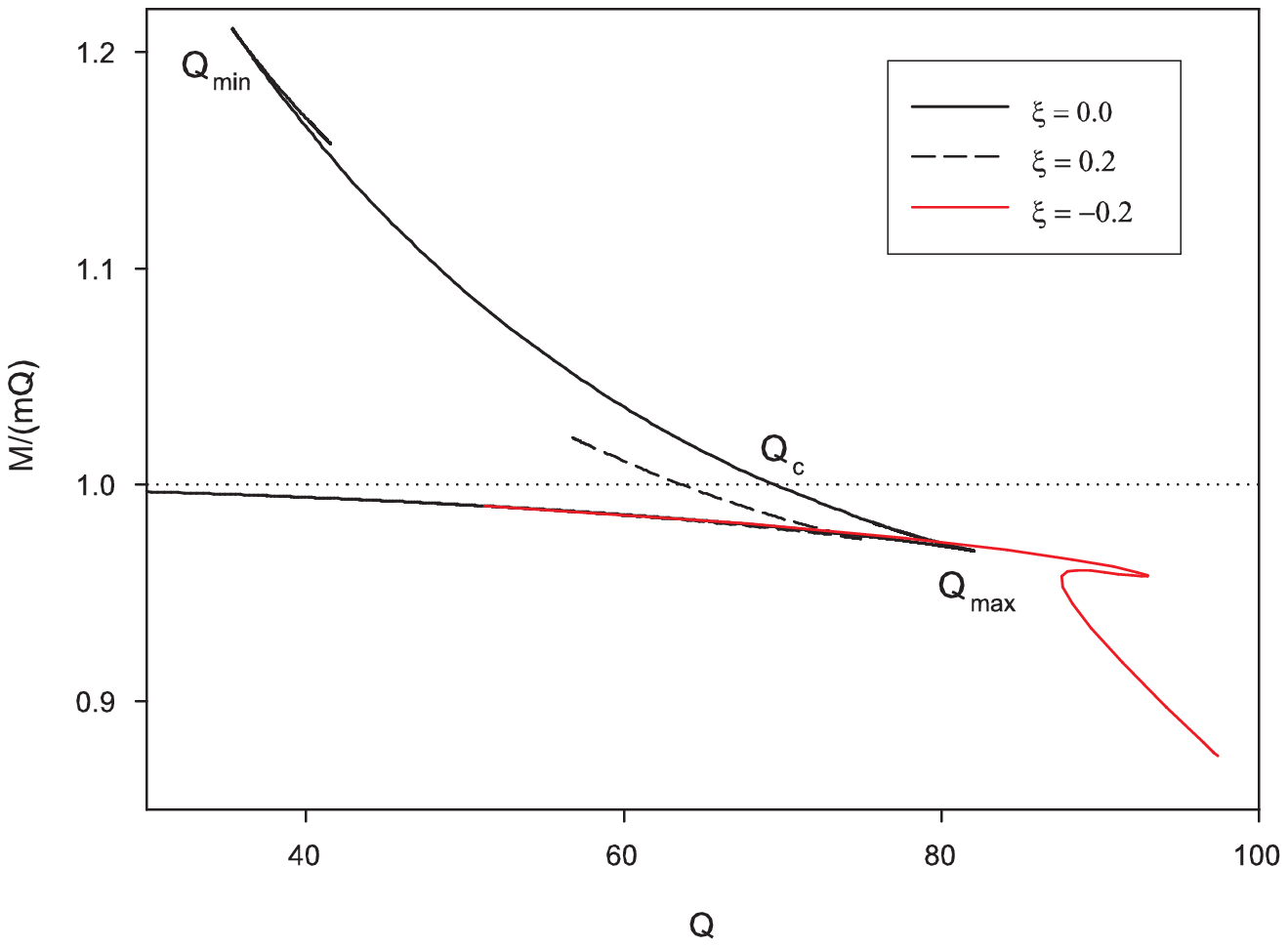}}
{\label{fig_1_bis_r}\includegraphics[width=8cm]{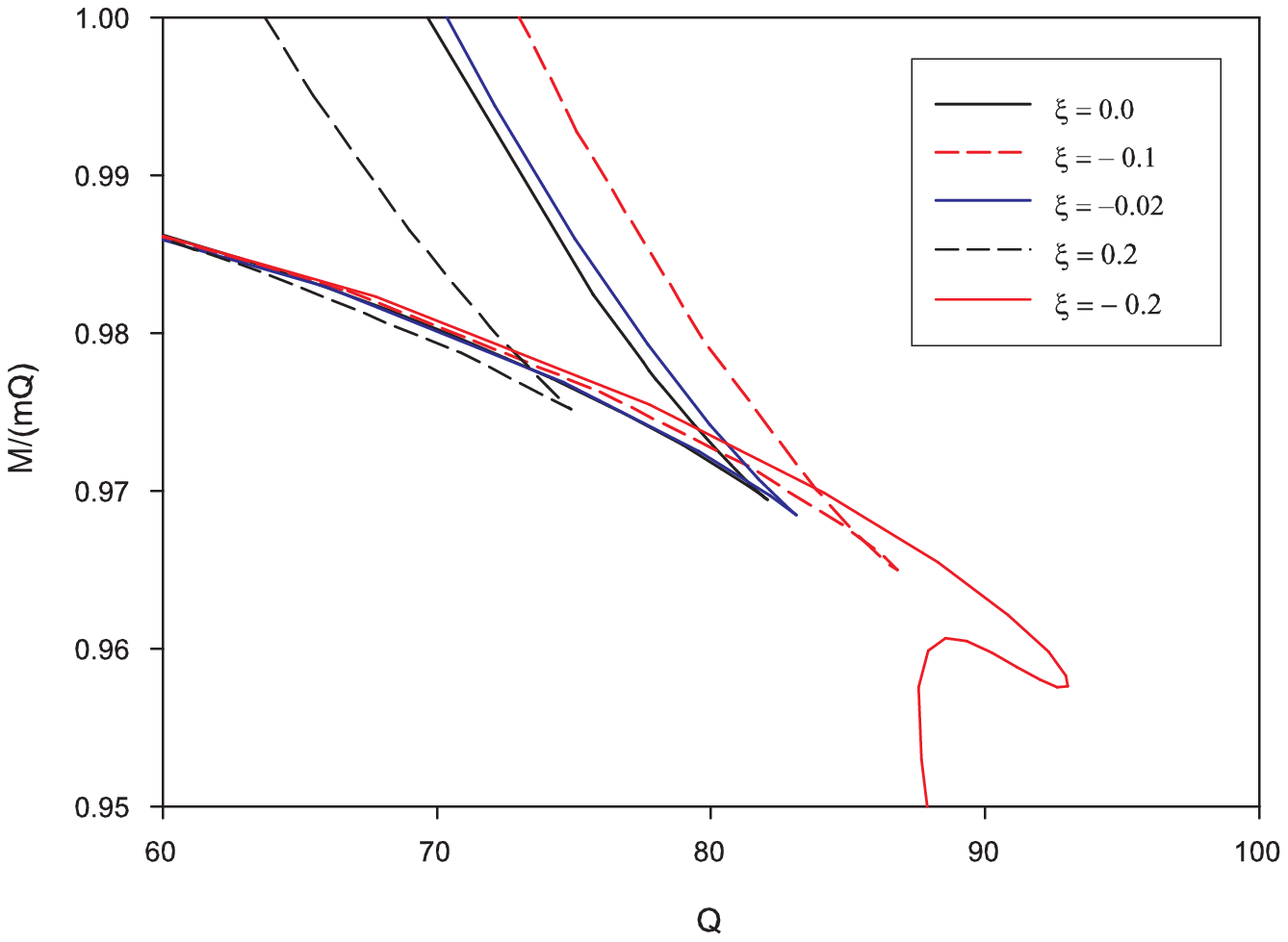}}
\end{center}
\caption{The ratio $M/Q$ as a function of $Q$ for several values of $\xi$.
\label{fig_1_bis}
}
\end{figure}

In this section we comment on some properties of BSs when the nonminimal coupling is absent. For more details please see \cite{Liebling:2012fv}.
In this case, the  constant $\kappa$ can be rescaled in the scalar field, and the mass $m$ of the scalar field can be rescaled in the radial variable.  We can therefore set $\kappa=1$, $m=1$ without losing generality. BSs are then essentially characterized by the central value $\phi_0$ of the soliton. 
In particular, the numerical integration   determines the frequency of the scalar field $\omega$ as a function of $\phi_0$. In this paper, we discuss only the fundamental solutions where the function $\phi(r)$ has no nodes; a series of excited solutions presenting zeros of $\phi(r)$ exist as well.
In the limit $\phi_0 \to 0$, the vacuum solution is approached ($M=Q=0$) and this corresponds to $\omega \to 1$.
Increasing gradually the parameter $\phi_0$, it turns out that the frequency $\omega$ first reaches a minimal value $\omega_m \approx 0.7677$ and then oscillates around an asymptotic mean value $\omega_a \sim 0.8425$ (see left part of Fig. \ref{fig_1}).
In spite of the fact the the frequency $\omega$ does not characterize the solutions uniquely,
it is common to display the mass $M$ and the charge $Q$ as functions of this parameter.  
Due to the oscillations, these plots currently present the form of  spirals as seen in the right part of Fig. \ref{fig_1}.\\
The three symbols bullet,  triangle and square symbolize the  special values where the charge $Q$ reaches its absolute maximum and minimum ($Q_{max}, Q_{min}$ and  
 $Q_c$ where $M = m Q_c$); these values play a role in the discussion of stability.
Completing Fig.\ref{fig_1}, we show on the left panel of Fig.\ref{fig_1_new} the dependance 
of the mass $M$ and of the charge $Q$
as functions of the central value of the scalar field $\phi_0$. The three exceptional values
$Q_{max}, Q_c, Q_{min}$ refer to the minimal case $\xi=0$ and corresponds, respectively, to $\phi_0 \sim 0.6, 1.37, 2.6$.
On the right panel of the figure, the dependance of $Q,M$ on the radius $R$ are reported.
Referring to the argument of stability invoked above, it
turns out  that the condition $M/Q < 1$ is fulfilled only for the small values of $\phi_0$, 
typically for $\phi_0 \leq 1.25$.
The value $\phi_0 = 1.37$ corresponds to $M=Q=Q_c \approx 67$, as seen in Fig. \ref{fig_1_bis}.
The plot of the ratio $M/Q$ as a function of $Q$ reveals the occurrence of at least three branches joining at spikes.
For later convenience let us call the branch connected to the vacuum (i.e. with $M=Q=0$)
the main branch and the other branches as the second, third branch and so on.
The  spike connecting the main and the second branches  corresponds to the maximal value of the charge,
say $Q=Q_{max}$. 
We find it for $Q_{max} \approx 82$, $M \approx 79.5$, $\omega \sim 0.85$, $\phi_0 = 0.6$;
it  belongs in the  domain of classical stability. The second spike connects the second and third branches 
and corresponds to a local minimum of $Q$, say $Q=Q_{min}$. We find
$Q_{min} \approx 36.00$, $M \approx 43.0$. This second spike 
 belongs to a region where the solutions are unstable. 
On the second branch, only the BSs  corresponding to $Q_c \leq Q \leq Q_{max}$ 
are classically stable. 

%

%
\subsection{Mini-boson stars with $\xi\neq 0$}
We  now discuss how the spectrum of the BSs is affected by the inclusion of the nonminimal coupling, i.e. for $\xi \neq 0$.
The classical equations now depend on two nontrivial parameters $\omega$ and $\xi$.
As expected by a continuity argument, integrating the field equations
for a fixed value of $\phi_0$, the minimally coupled BSs (i.e. with $\xi = 0$) 
can be  continuously deformed by increasing  (or decreasing) gradually the  coupling parameter $\xi$. \\

Let us first  discuss the case $\xi \neq 0$. Similarly to the case $\xi=0$,
a branch of BSs can then be constructed by increasing the parameter $\phi_0$. 
 This leads to
families of solutions  characterized by the frequency $\omega$, the charge $Q$ and the mass $M$.
In Fig. \ref{fig_1} we present some data  corresponding to different values of $\xi$ together with the case $\xi = 0$.
For $\xi \neq 0$, the curves  stop at some critical values of $\phi_0$;
the numerical integration indeed becomes problematic at some stage for high values of $\phi_0$. 
Our numerical results suggest that the critical phenomena limiting the solutions for positive and negative values of $\xi$ have different origins~:
\begin{itemize}
\item For {\bf  positive} values of $\xi$ the solutions cannot be constructed for large values of $\phi_0$ because the determinant $\Delta$ approaches  zero at a critical value of the parameter $\phi_0$, 
say $\phi_0 = \phi_{0,max}$. For example for $\xi = 0.2$, we find $\phi_{0,max} \approx 1.15$.
\item For {\bf  negative} values of $\xi$, the situation is different~: $\Delta$ decreases monotonically but not reaching zero while $\phi_0$ increases.   
\end{itemize}

One of the main effects of the nonminimal coupling is then to
limit the possible values of the central value $\phi_0$ of the boson
field. In particular setting $|\xi|> 0$ has the tendency to ``unwind" the spiral curves $M(\omega)$ 
as seen in Fig. \ref{fig_1}. Qualitatively, this resembles the effects of the Gauss-Bonnet interaction in the pattern of higher dimensional BSs in Einstein-Gauss-Bonnet gravity.
These solutions have been studied in \cite{Hartmann:2013tca} where it was shown that the origin of the critical phenomenon is related to the occurrence of a singularity of the metric at the origin.
In the present case, the geometry remains regular in the critical limit, instead the system of equations becomes singular when the determinant $\Delta$ approaches to zero. 

Before reexamining this phenomenon with a different point of view, let us discuss  the 
effects of the nonminimal coupling on the classical stability of the BSs. 
For small values of $|\xi|$, the plot of the ratio $M/Q$ as a function of $Q$ 
generally presents two branches joining in a spike at, say $Q=Q_{max}$
(see Fig. \ref{fig_1_bis}). 
The main branch  is stable all long. On the other hand a piece of the second branch is stable 
for $Q_{c} \leq Q \leq Q_{max}$ where we define $Q_{c}$ as the value of the charge where $M/Q = 1$.
For $Q \leq Q_c$, the solutions of the second branch are unstable.
Both values $Q_{c}, Q_{max}$ increase while $\xi$ decreases.
This scenario holds true for small enough values of $|\xi|$. 
Interestingly, for $\xi < - 0.15$ the pattern changes: Both the main and second branches are classically stable. 
Hence, negative values of  the nonminimal coupling  have the tendency to enhance the stability of the solutions.  
\begin{figure}[h]
\begin{center}
{\label{fig_3_l}\includegraphics[width=8cm]{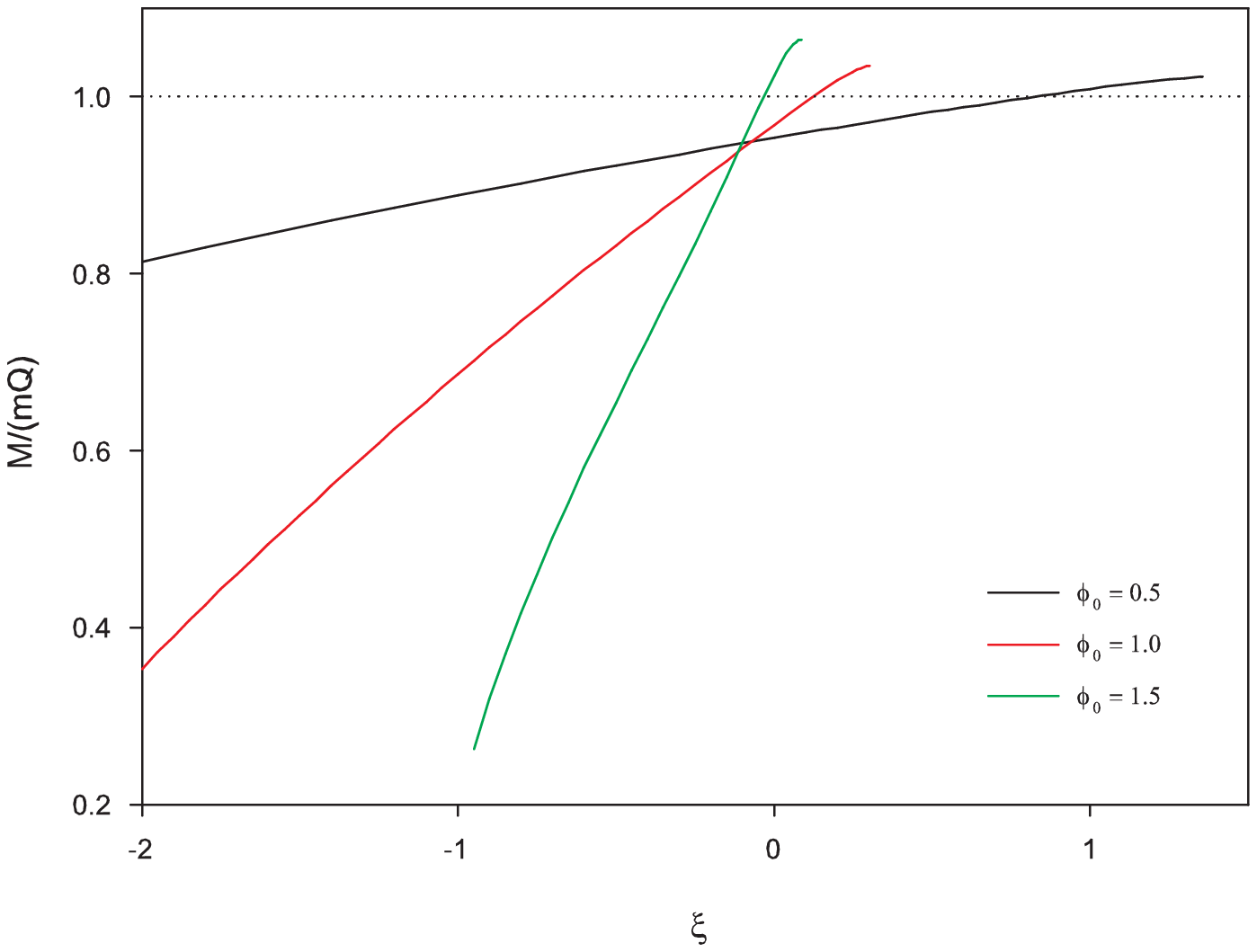}}
{\label{fig_3_r}\includegraphics[width=8cm]{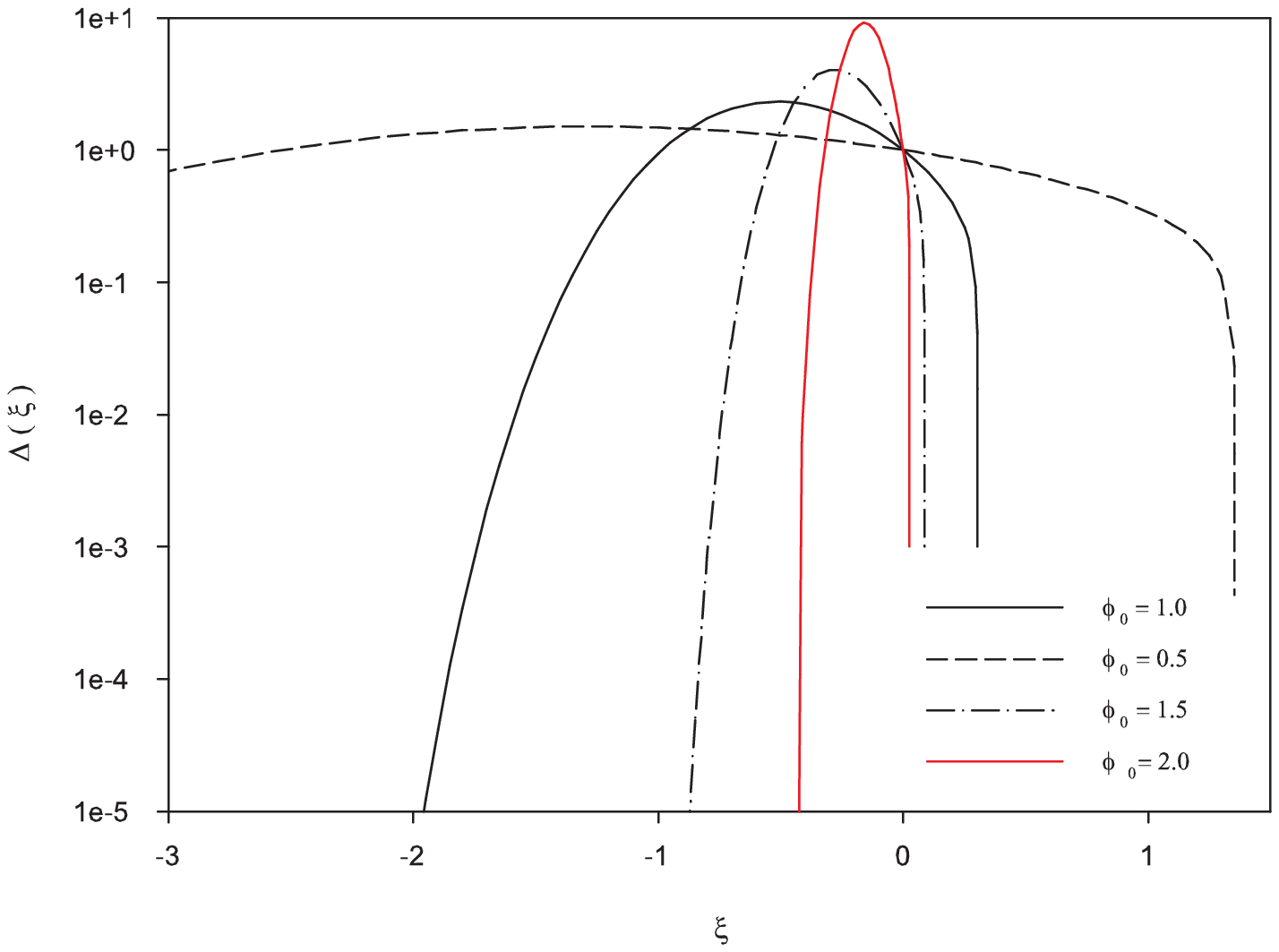}}
\end{center}
\caption{Left: Ratio $M/(mQ)$ as functions of $\xi$ for different values of $\phi_0$.
Right: Discriminant $\Delta$ as function of $\xi$  for several values of $\phi_0$.
\label{fig_3}
}
\end{figure}

To complete the discussion we study
 how solutions corresponding to a particular  central value $\phi_0$ 
are affected by the nonminimal coupling.
The results are the object of Figs. \ref{fig_3}. 
 The ratio $M/(mQ)$ is reported as a function of $\xi$ for three values of $\phi_0$ on the left panel:
 It shows that the ratio increases monotonically with $\xi$.
 As noticed already, the lump is more bounded. For negative values of $\xi$      
the critical phenomenon limiting the BSs for $|\xi| \neq 0$ is revealed on the right side of Fig \ref{fig_3}. 
  We see that the determinant $\Delta(\xi)$ suddenly approaches to zero for a {\it positive} critical value of $\xi$ (this value, depends of course on $\phi_0$).
In contrast, for $\xi < 0$, the value $\Delta$ regularly decreases to zero, although not 
reaching $\Delta=0$, while decreasing $\xi$. 
The numerical difficulties occur typically when $\Delta$ becomes of the order of the tolerance
imposed for the numerical integrator. We manage to construct robust solutions up to $\Delta \sim 10^{-8}$. 
For large values of $\phi_0$ (typically $\phi_0 \geq 2$) the following features, 
illustrated by the red curve in Fig. \ref{fig_3}, should be stressed 
\begin{itemize}
\item The value $\Delta$ becomes very sensitive to $\xi$ 
\item The interval of $\xi$ where the solutions exist decreases.  
\end{itemize}
These constitute the sources of  the numerical difficulties.


\section{Self-interacting solutions}\label{sec4}
\subsection{$\xi=0$ case}
We now discuss the effects of the self-interaction of the scalar field on the solutions.
As stated above, we choose  the particular potential
\be
          V_6(|\Phi|) = m^2 |\Phi|^2 ( |\Phi|^2 - 1 )^2
\ee 
which possesses a nontrivial vacuum manifold: $|\Phi|= 0$ and $|\Phi|=1$. 
Many properties of BSs in this potential
(including also the effect of an electric charge) have been discussed in \cite{Brihaye:2015veu}. 
Perhaps one of the main properties is that the BSs
can be continued to the nongravitating limit $\kappa = 0$, constituting a family of Q-ball solutions labeled by $\omega$. 
The self-interaction due to the potential confers very specific features to the Q-balls, 
some of which are shown in Fig. \ref{v6} (dashed lines)~: 
\begin{itemize}
\item The solutions exist up to a maximal value of $\phi_0$.
\item The solutions exist for arbitrarily small values of $\omega$. The limit $\omega \to 0$
corresponds to $\phi_0 \to 1$; the profile of the scalar field approaches 
a step function with $\phi(r) \sim 1$ for $r < R$ and with $\phi(r) \sim 0$ for $r > R$,
so that the boson field is essentially concentrated in a sphere of radius $R$.
This corresponds to the so-called ``thin-wall limit'';   
the mass, the charge and the radius $R$ diverge while $\omega$ approaches zero.
\item In the limit $\phi_0 \to 0$ the matter field approaches uniformly the vacuum configuration $\phi(r)=0$
although the mass and the charge remain finite, forming a ``mass gap''. 
This is denoted by $Y$ on the right side of Fig.\ref{v6}.
\end{itemize}
The coupling to gravity has the effect to regularize the  Q-balls configurations. This is  
shown in Fig. \ref{v6} where the data corresponding to $\kappa =0.1$ (we set $\xi = 0$ in this section)
  is reported by means of the
solid lines. In contrast with Q-balls, the following features hold~:
\begin{itemize}
\item BSs  exist  for large  values of $\phi_0$. The mass and charge  remain finite and bounded.
\item There is a minimal value of $\omega$. The minimal value depend on the constant $\kappa$. 
\item In the limit $\phi_0 \to 0$ the matter field approaches uniformly the vacuum configuration $\phi(r)=0$. 
The mass and the charge converge to zero.
\end{itemize}
\begin{figure}[ht!]
\begin{center}
{\label{v6}\includegraphics[width=8cm]{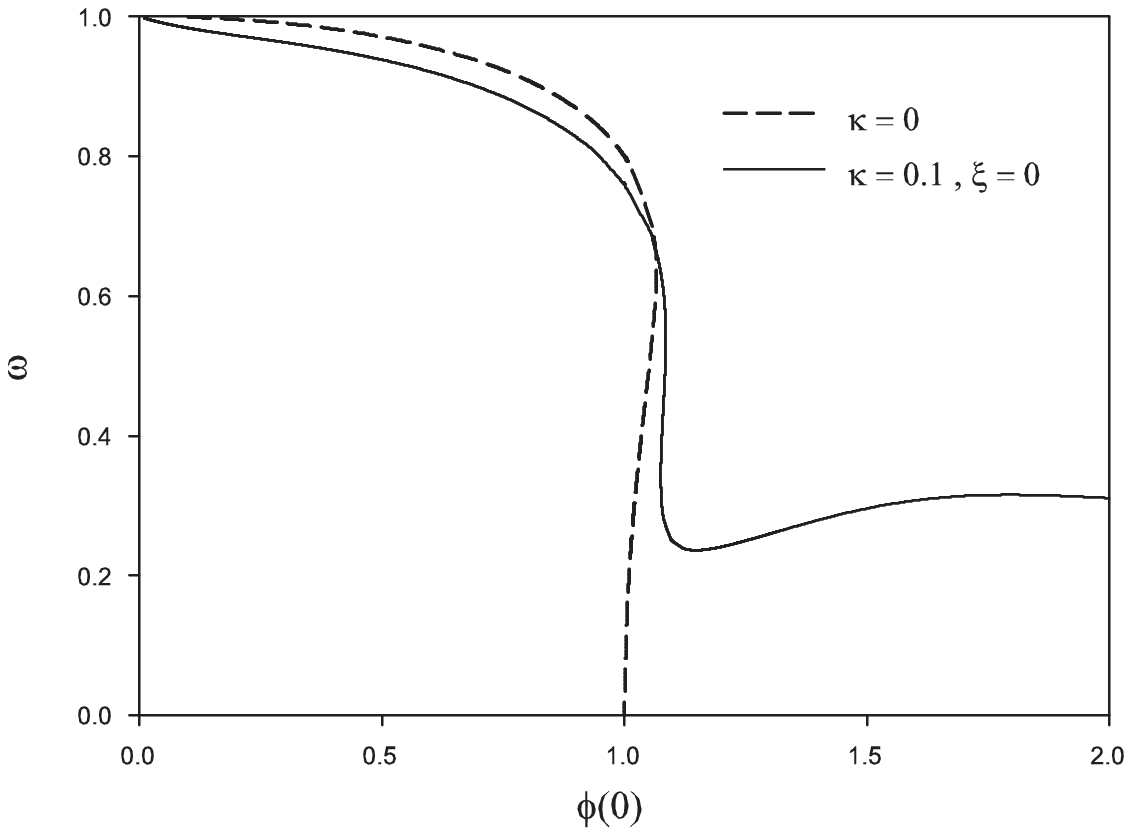}}
{\label{hyper_r}\includegraphics[width=8cm]{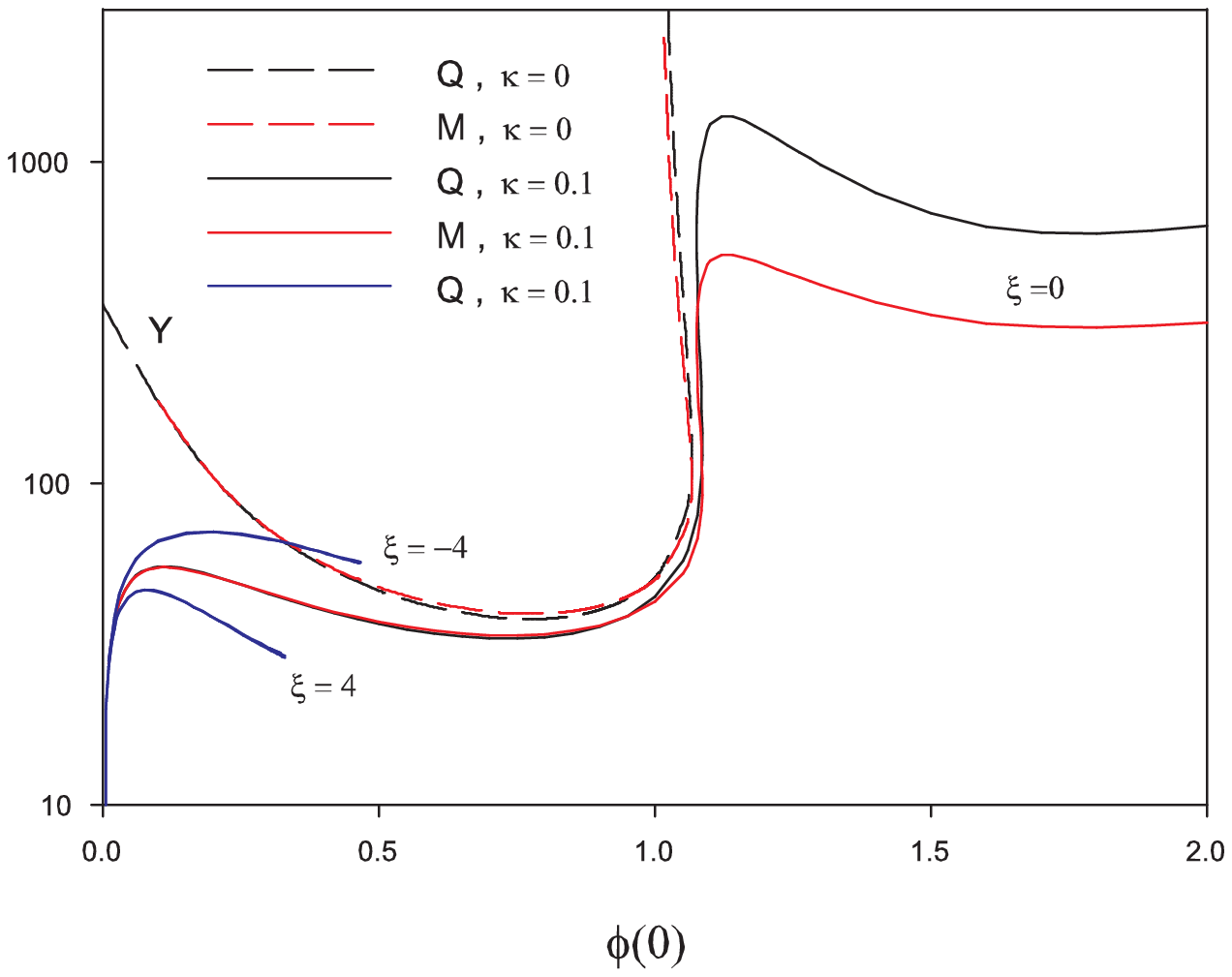}}
\end{center}
\caption{Left: The dependence of $\omega $ on $\phi_0$ for Q-balls (dashed) and BSs (solid). 
Right: The mass, charge dependance of $\phi_0$. 
\label{v6}
}
\end{figure}

The classical stability of self-interacting Q-balls and (minimal-coupled) BSs 
can be read from the $M/Q$ plot provided in Fig.\ref{fig_5}. 
The curve corresponding to BSs is the black-solid line. It shows the occurrence of three
branches joining in two spikes (labeled $A$ and $B$ in the figure) and forming a curve with the shape of a butterfly. 
The main branch, connected to the vacuum and terminating at A,  corresponds to a set of stable solutions.
The intermediate branch $A-B$ is essentially unstable (only on a small fraction of it the solutions are stable).
The third branch terminating at $B$ is stable in its part corresponding to large values of $\phi_0$.
\begin{figure}[ht!]
\begin{center}
{\label{fig_5}\includegraphics[width=14cm]{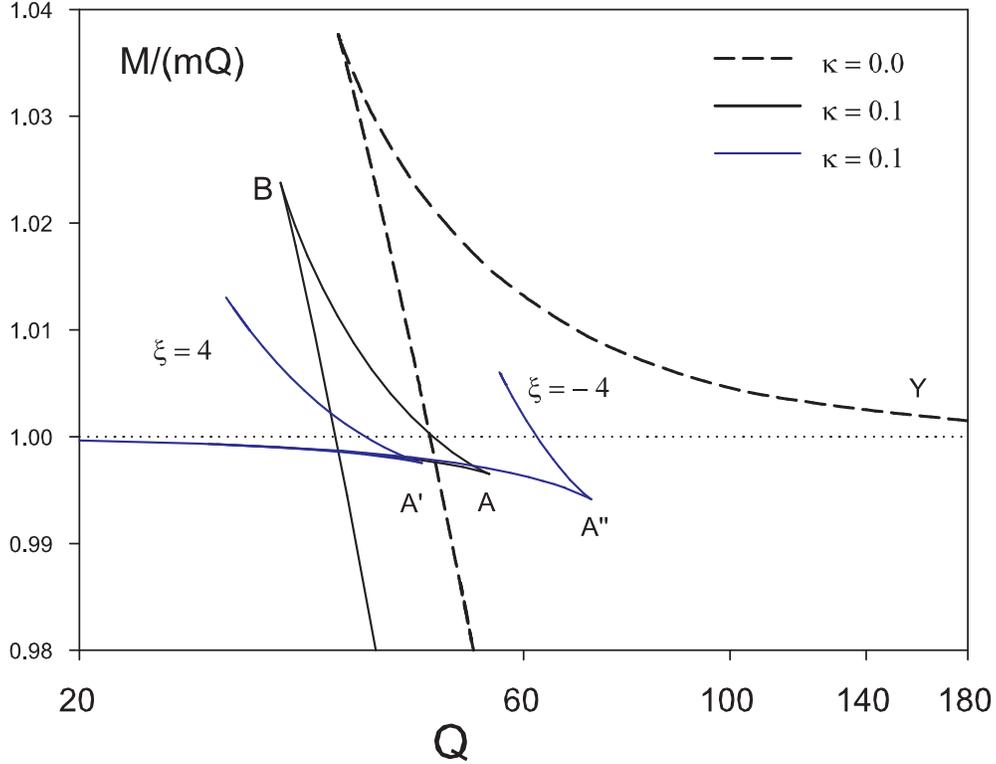}}
\end{center}
\caption{ The ratio $M/(mQ)$ as function of $Q$ for different self-interacting solutions.
\label{fig_5}
}
\end{figure}

\subsection{$\xi\neq0$ case}
We now discuss the influence of the nonminimal coupling to the solutions. 
For definiteness we set $\kappa = 0.1$ in our numerical construction.
Following the same lines
as in the previous section, we analyzed the deformation of the BSs for $\xi \neq 0$. 
As expected, it turns out that the nonminimal coupling 
reduces considerably the domain of existence of the BSs: For both signs of $\xi$ the solutions 
exist only for small enough values of $\phi_0$. In particular
the quantity $\Delta(\xi)$ approaches zero while increasing  the value of $\phi_0$ of the solution,
leading to a maximal value, say $\phi_{0,max}$. The precise determination of $\phi_{0,max}(\xi)$ is beyond the scope of this paper, but our numerical results demonstrate that it is monotonically decreasing
while $|\xi|$ increases.
The data corresponding to $\xi = \pm 4$ is shown in Fig. \ref{v6} 
(the blue lines for the charge $Q$, the curve for the corresponding mass is very close and not reported). 

Remembering that the self-interaction allows for solutions to exist in the absence of normal gravity, 
the question arises naturally whether solitons  
could interact with gravity through the nonminimal derivative term only, i.e. with $\alpha = 0, \xi = 1$. 
We therefore put some emphasis on solutions with $\alpha/|\xi| \ll 1$.
Our numerical results strongly suggest that the 
standard BSs  do not survive in the limit $\alpha/ |\xi| \to 0$,
their domain being too restricted by the condition $\Delta(\xi) > 0$. 
It is possible, however, that new types of solitons exist on a domain of the parameter space
connected to  $\alpha = 0, \xi = 1$.
This would constitute a bosonic counterpart of the neutron stars obtained
in \cite{Cisterna:2015yla}.
So far, we failed to construct such  solutions numerically.

Let us finally comment on how the stability pattern is affected for $\xi \neq 0$.
Due to the reduction of the domain of the solutions, 
the ``butterfly'' curve occurring for $\xi = 0$ is  progressively reduced as well.
 For the cases $\xi =  4$ and $\xi = -4$, chosen for Fig. \ref{fig_5}, only two of the three branches remain;
 they are joining at points $A'$ and $A''$ respectively. 
 The solutions on the branch joining to the vacuum are stable, irrespectively of the sign of $\xi$. Negative values of $\xi$
  allow for stable solutions with higher values of the charge $Q$ and of the  energy binding $M/Q-1$.

\section{Final remarks}\label{sec5}
In this work we have constructed BS configurations for STT possessing a nonstandard kinetic term coupled through the Einstein tensor. This particular coupling is contained in the most general STT with second order equations of motion for a single new scalar degree of freedom, the so called Horndeski theory. \\
Due to the fact that we are dealing with a complex scalar field, instead of Horndeski gravity, our model is embedded in its biscalar extension, namely, in the context of the most general STT with second order equations of motion, constructed with a single massless metric tensor and with two real scalar field degrees of freedom. In this scenario BSs are supported by new degrees of freedom and not by external matter sources. It is important to stress that along with the new degrees of freedom also external matter fields may be included. In the context of STT, in \cite{Alcubierre:2010ea, Ruiz:2012jt, Kleihaus:2015iea} the authors have tackled this problem showing that the phenomenon of spontaneous scalarization originally predicted for neutron stars, can also occur for BSs. Moreover, we are not considering here any kind of interaction between external fields and new scalar degrees of freedom.\\
We have analyzed the existence of mini-BSs configurations (where only a mass term is considered) and of self-interacting BSs where the self interaction possesses a sixth order potential which can be written, for specific values of the involved couplings, as a positive definite potential presenting two degenerate local vacua. In both cases we have shown that the determinant of our system of equations (\ref{system}) plays a fundamental role in the pattern of solutions. Indeed, when this determinant approaches to zero no solutions can be obtained. In practice we have seen that the pole of this determinant is always located at the origin, luring as to define the control parameter $\Delta$ (\ref{control}) in order to look for nonsingular solutions.\\
Mini-BS solutions exist for both, positive and negative values of the nonminimal rescaled parameter $\xi$.
For $\xi>0$ the solutions cannot be obtained when the central value $\phi_0$ exceeds some maximal value $\phi_0 = \phi_{0,max}$, for which our function $\Delta$ goes to zero. On the other hand, the $\xi<0$ case is different. Here, $\Delta$ shows a monotonically decreasing behavior when $\phi_0$ is increased without reaching the conflictual point $\Delta=0$; nevertheless complications arise when this function approaches to values of the same order than the tolerance imposed by the numerical integrator. The $\xi$ negative branch also shows a tendency to enhance the stability of the solutions.\\
For the self-interacting solutions considered here, the situation is similar to the mass term case. The pattern of solutions is harshly constrained and exists for a limited branch of values of $\phi_0$, later that indeed depend on $\xi$. For this particular case we have also investigated the existence of configurations supported only by the presence of the nonminimal kinetic coupling, this means for the $\alpha/|\xi|\rightarrow0$ case. Our results suggest that BSs do not survive in this case. It would be interesting to circumvent this problem in order to construct the bosonic counterpart of the neutron stars constructed in \cite{Cisterna:2015yla} and make qualitative comparisons. We leave this for future work. 

\section*{Acknowledgments}

A.C. would like to express his gratitude to the Theoretical Physics Department of the University of Geneva for its kind hospitality during the final stage of this work. Y.B. and A.C. would like to thank M. Hassaine, B. Hartmann, M. Rinaldi, F. Canfora and T. Delsate for interesting discussions and comments. A.C.'s work is supported by FONDECYT project N\textordmasculine3150157. The work of C.E. is supported by CONICYT and Centro de Estudios Cient\'{\i}ficos (CECs)
funded by the Chilean Government through the Centers of Excellence Base
Financing Program of CONICYT.
\section{Appendix: Field equations}
Our field equations can be cast in the following matrix form
\begin{equation}
AB=C
\end{equation}
where, before the rescaling made in Sec. \ref{rescaling}, we have defined
\begin{equation}
A= 
 \begin{bmatrix}
A_{11} & A_{12} & A_{13}  \\
A_{21} & A_{22} & A_{23}  \\
A_{31} & A_{32} & A_{33} 
\end{bmatrix}
\end{equation}
\begin{equation}
B= 
 \begin{bmatrix}
F''  \\
G''   \\
\phi''  
\end{bmatrix}
\end{equation}
\begin{equation}
C= 
 \begin{bmatrix}
K_1(F,F',G,G',\phi,\phi',\tilde{\omega})  \\
K_2(F,F',G,G',\phi,\phi',\tilde{\omega})   \\
K_3(F,F',G,G',\phi,\phi',\tilde{\omega})  
\end{bmatrix}
\end{equation}
with
\begin{equation*}
A_{11}=\frac{r}{4}\left(\frac{-6\tilde{\omega}^2\phi^2\eta G^3Fr+4\kappa F^2G^3r+2rG^2F^3\phi'^2\eta}{G^{5/2}F^4}\right)
\end{equation*}

\begin{equation*}
A_{12}=\left(\frac{r^2\tilde{\omega}^2\phi^2\eta}{\sqrt{G}F^2}\right)
\end{equation*}

\begin{equation*}
A_{13}=\frac{r}{4}\left(\frac{4rG^2F^3\phi'\eta F'+8rG^3F^2\tilde{\omega}^2\phi\eta}{G^{5/2}F^4}\right)
\end{equation*}

\begin{equation*}
A_{21}=\left(\frac{r^2\tilde{\omega}^2\phi^2\eta}{\sqrt{G}F^2}\right)=A_{12}
\end{equation*}

\begin{equation*}
A_{22}=-\frac{1}{8}\left(\frac{8r^2G^2\kappa F^3+4r^2G^2\tilde{\omega}^2\phi^2\eta F^2+4r^2G\phi'^2\eta F^4}{G^{7/2}F^3}\right)
\end{equation*}

\begin{equation*}
A_{23}=-\frac{1}{8}\left(\frac{16rG^2\phi'\eta F^4+16r^2G^3\tilde{\omega}^2\phi\eta F^2+8r^2G\phi'\eta F^4G'}{G^{7/2}F^3}\right)
\end{equation*}

\begin{equation*}
A_{31}=\frac{r}{4}\left(\frac{4rG^2F^3\phi'\eta F'+8rG^3F^2\tilde{\omega}^2\phi\eta}{G^{5/2}F^4}\right)=A_{13}
\end{equation*}

\begin{equation*}
A_{32}=-\frac{1}{8}\left(\frac{16rG^2\phi'\eta F^4+16r^2G^3\tilde{\omega}^2\phi\eta F^2+8r^2G\phi'\eta F^4G'}{G^{7/2}F^3}\right)=A_{23}
\end{equation*}

\begin{equation*}
\begin{split}
A_{33}=&-\frac{1}{4}\left(\frac{-8r^2F^3G^4\alpha+2r^2F^4G\eta G'^2-2r^2F^2G^3\eta F'^2+8rF^4G^2\eta G'}{G^{7/2}F^3}\right)\\
\end{split}
\end{equation*}

\begin{equation*}
\begin{split}
K_{1}=&-\frac{r}{4G^{5/2}F^4}(-4\lambda_{2}\phi^4F^2G^4r+4\lambda_{3}\phi^6F^2G^4r\\
&+2\kappa F^2G^2G'rF'-8\tilde{\omega}^2\phi^2\alpha FG^4r+9\tilde{\omega}^2\phi^2\eta G^3F'^2r\\
&-12\tilde{\omega}^2\phi^2\eta G^3F'F+8\tilde{\omega}^2\phi^2\eta G^2G'F^2-\phi'^2\eta F^2G^2F'^2r\\
&+\phi'^2\eta F^4G'^2r-4\kappa FG^3F'^2r+16G^3F^2\tilde{\omega}^2\phi\eta \phi'\\
&+4\lambda_{1}\phi^2F^2G^4r-3\tilde{\omega}^2\phi^2\eta G^2G'FrF'+4G^2F^3\phi'^2\eta F'\\
&+4\phi'^2\eta F^4G'G+8\kappa F^2G^3F'+8rG^3F^2\tilde{\omega}^2\phi'^2\eta\\
&+4rG^2F^2\tilde{\omega}^2\phi\eta G'\phi'-rGF^3\phi'^2\eta F'G'\\
&-12rG^3F\tilde{\omega}^2\phi\eta F'\phi'-3\tilde{\omega}^2\phi^2\eta GG'^2F^2 r)
\end{split}
\end{equation*}

\begin{equation*}
\begin{split}
K_{2}=&\frac{1}{8G^{7/2}F^3}(8rG\phi'^2\eta F^4G'-5r^2\phi'^2\eta F^4G'^2-12r^2G^4\tilde{\omega}^2\phi^2\alpha F\\
&+13r^2G^3\tilde{\omega}^2\phi^2\eta F'^2-r^2G^2\phi'^2\eta F^2F'^2+4r^2G\phi'^2\eta F^3G'F'\\
&+8rG^2\phi'^2\eta F^3F'+16r^2G^3\tilde{\omega}^2\phi'^2\eta F^2-6r^2F^3GG'^2\kappa\\
&+16rG^2\kappa F^3G'+12r^2G^4\lambda_{1}\phi^2F^2-12r^2G^4\lambda_{2}\phi^4F^2\\
&+4r^2G^3\phi'^2\alpha F^3+8F^4G^2\phi'^2\eta-3r^2F^2GG'^2\tilde{\omega}^2\phi^2\eta\\
&+32rG^3\tilde{\omega}^2\phi\eta F^2\phi'-16rG^3\tilde{\omega}^2\phi^2\eta F'F+8rG^2\tilde{\omega}^2\phi^2\eta G'F^2\\
&+8r^2G^2\tilde{\omega}^2\phi\eta G'F^2\phi'-4r^2G^2\tilde{\omega}^2\phi^2\eta G'FF'\\
&-24r^2G^3\tilde{\omega}^2\phi\eta FF'\phi'+2r^2G^3\kappa FF'^2+12r^2G^4\lambda_{3}\phi^6F^2)
\end{split}
\end{equation*}

\begin{equation*}
\begin{split}
K_{3}=&\frac{1}{4G^{7/2}F^3}(r^2\phi'F^2G'\eta G^2F'^2+2r^2F^3G\phi'\eta G'^2F'\\
&-4r^2F^3G^3\phi'\alpha G'-8\phi'F^4G\eta G'^2r-8r^2\phi G^5\tilde{\omega}^2\alpha F\\&+10r^2\phi G^4\tilde{\omega}^2\eta F'^2
-4\phi'F^2G^3\eta F'^2r+16r\phi G^3\tilde{\omega}^2\eta G'F^2\\
&+2r^2FG^3\phi'\eta F'^3+8r^2\phi G^5\lambda_{1}F^2-6r^2\phi G^2\tilde{\omega}^2\eta G'^2F^2\\
&+24r^2\phi^5G^5\lambda_{3}F^2-5r^2\phi'F^4G'^3\eta+8\phi'F^4G^2\eta G'\\
&-4r^2\phi G^3\tilde{\omega}^2\eta G'FF'-16r\phi G^4\tilde{\omega}^2\eta F'F-16\phi'F^3G^4\alpha r\\
&+8rF^3G^2\phi'\eta G'F'-16r^2\phi^3G^5\lambda_{2}F^2)
\end{split}
\end{equation*}


\begin{thebibliography}{99}

  \bibitem {BD}C.~Brans and R.~H.~Dicke,
Phys.\ Rev.\ \textbf{124}, 925 (1961).

\bibitem {Horndeski}G.~W.~Horndeski, Int.\ J.\ Theor.\ Phys.\ \textbf{10}, 363 (1974).

\bibitem{Nicolis} 
A.~Nicolis, R.~Rattazzi and E.~Trincherini,
  Phys.\ Rev.\ D {\bf 79}, 064036 (2009).
  
 \bibitem{DGP} G.R~ Dvali, G.~Gabadadze, and M.~Porrati, Phys.\ Lett.\ \textbf{B} 485, 208 (2000). 

\bibitem{Deffayet:2000uy} 
  C.~Deffayet,
  Phys.\ Lett.\ B {\bf 502}, 199 (2001).
 
\bibitem{Lue:2005ya} 
  A.~Lue,
  Phys.\ Rept.\  {\bf 423}, 1 (2006).
 
\bibitem{Deffayet:2001pu} 
  C.~Deffayet, G.~R.~Dvali and G.~Gabadadze,
  Phys.\ Rev.\ D {\bf 65}, 044023 (2002).

 
\bibitem{Babichev:2013usa} 
  E.~Babichev and C.~Deffayet,
  Class.\ Quant.\ Grav.\  {\bf 30}, 184001 (2013).

 \bibitem{galileon1}C.~Deffayet, G.~Esposito-Far\'ese, and A.~Vikman, Phys.\ Rev.\ \textbf{D} 79, 084003 (2009).

\bibitem{Deffayet:2009mn} 
  C.~Deffayet, S.~Deser and G.~Esposito-Farese,
  Phys.\ Rev.\ D {\bf 80}, 064015 (2009).
  
\bibitem{Kobayashi:2011nu} 
  T.~Kobayashi, M.~Yamaguchi and J.~Yokoyama,
  Prog.\ Theor.\ Phys.\  {\bf 126}, 511 (2011).
  
   \bibitem{fabfour}
 C.~Charmousis, E.~J.~Copeland, A.~Padilla and P.~M.~Saffin,
  Phys.\  Rev.\ Lett.\  {\bf 108} 051101 (2012).

\bibitem{Weinberg:1988cp} 
  S.~Weinberg,
  Rev.\ Mod.\ Phys.\  {\bf 61}, 1 (1989).
  
\bibitem{Germani:2010gm} 
  C.~Germani and A.~Kehagias,
  Phys.\ Rev.\ Lett.\  {\bf 105}, 011302 (2010).

\bibitem{Amendola:1993uh} 
  L.~Amendola,
  Phys.\ Lett.\ B {\bf 301}, 175 (1993).

\bibitem{Sushkov:2009hk} 
  S.~V.~Sushkov,
  Phys.\ Rev.\ D {\bf 80}, 103505 (2009).

\bibitem{Myrzakulov:2015ysa} 
  R.~Myrzakulov and L.~Sebastiani,
  arXiv:1512.00402 [gr-qc].

\bibitem{Gumjudpai:2015vio} 
  B.~Gumjudpai and P.~Rangdee,
  Gen.\ Rel.\ Grav.\  {\bf 47}, no. 11, 140 (2015).
  
  
   \bibitem{namur}
  J.~P.~Bruneton, M.~Rinaldi, A.~Kanfon, A.~Hees, S.~Schlogel and A.~Fuzfa,
  Adv.\ Astron.\  {\bf 2012} 430694 (2012) .
  
    \bibitem{pert}
  A.~De Felice, T.~Kobayashi and S.~Tsujikawa,
  Phys.\ Lett.\ B {\bf 706} 123 (2011). 
  
   F.~Piazza and F.~Vernizzi,
  Class.\ Quant.\ Grav.\  {\bf 30} 214007 (2013).
  
  A.~De Felice and S.~Tsujikawa,
  JCAP {\bf 1202} 007 (2012) .

  
\bibitem{Amendola:2012ky} 
  L.~Amendola, M.~Kunz, M.~Motta, I.~D.~Saltas and I.~Sawicki,
  Phys.\ Rev.\ D {\bf 87}, no. 2, 023501 (2013).
  
\bibitem{Motta:2013cwa} 
  M.~Motta, I.~Sawicki, I.~D.~Saltas, L.~Amendola and M.~Kunz,
 Phys.\ Rev.\ D {\bf 88}, no. 12, 124035 (2013).
  
  
  






 

 

 
 
 

  

  




  

  
 
\bibitem{Hui:2012qt} 
  L.~Hui and A.~Nicolis,
  Phys.\ Rev.\ Lett.\  {\bf 110}, 241104 (2013).

 
\bibitem{rinaldi}
 M.~Rinaldi,
  Phys.\ Rev.\ D {\bf 86} 084048 (2012).

\bibitem{Kolyvaris:2011fk} 
  T.~Kolyvaris, G.~Koutsoumbas, E.~Papantonopoulos and G.~Siopsis,
  Class.\ Quant.\ Grav.\  {\bf 29}, 205011 (2012).

\bibitem{charmousis1} E. Babichev and C. Charmousis, JHEP \textbf{1408}, 106 (2014).

\bibitem{adolfo2} A.~Anabal\'on, A.~Cisterna and J.~Oliva, Phys.\ Rev.\  D \textbf{89}, 084050 (2014).

\bibitem{Minam1} M. Minamitsuji, Phys. Rev. D \textbf{89}, 064017 (2014).

\bibitem{Kobayashi:2014eva} 
  T.~Kobayashi and N.~Tanahashi,
  PTEP {\bf 2014}, 73E02 (2014).

\bibitem{adolfo3} A.~Cisterna and C.~Erices, Phys.\ Rev.\ D \textbf{89}, 084038 (2014).


\bibitem{MK} M.~Bravo-Gaete and M.~Hassaine,
  Phys.\ Rev.\ D {\bf 89}, 104028 (2014).

\bibitem {minas} C. Charmousis, T. Kolyvaris, E. Papantonopoulos, and M. Tsoukalas, JHEP \textbf{07}, 085 (2014).

\bibitem{Ogawa} 
  H.~Ogawa, T.~Kobayashi and T.~Suyama,
  Phys.\ Rev.\ D {\bf 93}, 064078 (2016)

\bibitem{Maselli:2015yva} 
  A.~Maselli, H.~O.~Silva, M.~Minamitsuji and E.~Berti,
  Phys.\ Rev.\ D {\bf 92}, no. 10, 104049 (2015).
 


\bibitem{Cisterna:2015uya} 
  A.~Cisterna, M.~Cruz, T.~Delsate and J.~Saavedra,
  Phys.\ Rev.\ D {\bf 92}, no. 10, 104018 (2015).
    
\bibitem{Koutsoumbas:2015ekk} 
  G.~Koutsoumbas, K.~Ntrekis, E.~Papantonopoulos and M.~Tsoukalas,
  arXiv:1512.05934 [gr-qc].

  
\bibitem{Cisterna:2015yla} 
  A.~Cisterna, T.~Delsate and M.~Rinaldi,
  Phys.\ Rev.\ D {\bf 92}, no. 4, 044050 (2015).
  

\bibitem{Silva:2016smx} 
  H.~O.~Silva, A.~Maselli, M.~Minamitsuji and E.~Berti,
  arXiv:1602.05997 [gr-qc].
  
\bibitem{Cisterna:2016vdx} 
  A.~Cisterna, T.~Delsate, L.~Ducobu and M.~Rinaldi,
  arXiv:1602.06939 [gr-qc].

\bibitem{Demorest:2010bx} 
  P.~Demorest, T.~Pennucci, S.~Ransom, M.~Roberts and J.~Hessels,
  Nature {\bf 467}, 1081 (2010).


\bibitem{Maselli:2016gxk} 
  A.~Maselli, H.~O.~Silva, M.~Minamitsuji and E.~Berti,
  arXiv:1603.04876 [gr-qc].



\bibitem{Manton:2004tk} 
  N.~S.~Manton and P.~Sutcliffe, {\it Topological solitons}, Cambridge University Press (2004).
    
\bibitem{Kaup:1968zz} 
  D.~J.~Kaup,
  Phys.\ Rev.\  {\bf 172}, 1331 (1968).
    
\bibitem{Kusmartsev:1990cr} 
  F.~V.~Kusmartsev, E.~W.~Mielke and F.~E.~Schunck,
  Phys.\ Rev.\ D {\bf 43}, 3895 (1991).
    
    
\bibitem{Kleihaus:2011sx} 
  B.~Kleihaus, J.~Kunz and S.~Schneider,
  Phys.\ Rev.\ D {\bf 85}, 024045 (2012).
    
    
    
    
     \bibitem{Lee:1991ax}
  T.~D.~Lee and Y.~Pang,
  Phys.\ Rept.\  {\bf 221}  251 (1992).
 
  
\bibitem{Volkov:2002aj} 
  M.~S.~Volkov and E.~W\"ohnert,
  Phys.\ Rev.\ D {\bf 66}, 085003 (2002).
 
  
\bibitem{Kleihaus:2005me} 
  B.~Kleihaus, J.~Kunz and M.~List,
  Phys.\ Rev.\ D {\bf 72}, 064002 (2005).
  
  
\bibitem{Kusenko:1997zq} 
  A.~Kusenko,
  Phys.\ Lett.\ B {\bf 405}, 108 (1997).
  
\bibitem{Kusenko:1997ad} 
  A.~Kusenko,
  Phys.\ Lett.\ B {\bf 404}, 285 (1997).
  
  
\bibitem{Hartmann:2012gw}
  B.~Hartmann and J.~Riedel,
  Phys.\ Rev.\ D {\bf 87}, 044003 (2013).
    
\bibitem{Friedberg:1986tq} 
  R.~Friedberg, T.~D.~Lee and Y.~Pang,
  Phys.\ Rev.\ D {\bf 35}, 3658 (1987).
  
\bibitem{Jetzer:1991jr} 
  P.~Jetzer,
  Phys.\ Rept.\  {\bf 220}, 163 (1992).
  
  
\bibitem{Liddle:1993ha} 
  A.~R.~Liddle and M.~S.~Madsen,
  Int.\ J.\ Mod.\ Phys.\ D {\bf 01}, 101 (1992).
    
\bibitem{Kusenko:1997si} 
  A.~Kusenko and M.~E.~Shaposhnikov,
  Phys.\ Lett.\ B {\bf 418}, 46 (1998).
  
\bibitem{Eby:2015hsq} 
  J.~Eby, C.~Kouvaris, N.~G.~Nielsen and L.~C.~R.~Wijewardhana,
  JHEP {\bf 1602}, 028 (2016).
  
  \bibitem{shapiro}M.~Colpi, S.~L.~Shapiro, and I.~Wasserman, Phys. Rev. Lett. \textbf{57}, 2485 (1986).
  
    
    
\bibitem{Vincent:2015xta} 
  F.~H.~Vincent, Z.~Meliani, P.~Grandclement, E.~Gourgoulhon and O.~Straub,
  arXiv:1510.04170 [gr-qc].
    
\bibitem{Torres:2000dw} 
  D.~F.~Torres, S.~Capozziello and G.~Lambiase,
  Phys.\ Rev.\ D {\bf 62}, 104012 (2000).
  
\bibitem{Macedo:2013jja} 
  C.~F.~B.~Macedo, P.~Pani, V.~Cardoso and L.~C.~B.~Crispino,
  Phys.\ Rev.\ D {\bf 88}, no. 6, 064046 (2013).
  
\bibitem{Meliani:2015zta} 
  Z.~Meliani, F.~H.~Vincent, P.~Grandclément, E.~Gourgoulhon, R.~Monceau-Baroux and O.~Straub,
  Class.\ Quant.\ Grav.\  {\bf 32}, no. 23, 235022 (2015).
  
\bibitem{Cunha:2015yba} 
  P.~V.~P.~Cunha, C.~A.~R.~Herdeiro, E.~Radu and H.~F.~Runarsson,
  Phys.\ Rev.\ Lett.\  {\bf 115}, no. 21, 211102 (2015).
    
  \bibitem{Ghez}
A. M. Ghez, B. L. Klein, M. Morris , and E. E. Becklin, ApJ. {\bf 509}, 678 (1998).

\bibitem{TV}
D. Tsiklauri, and R. D. Viollier, ApJ. {\bf 500}, 591 (1998).  


\bibitem{RB} R.~Ruffini and S.~Bonazzola, Phys.\ Rev.\  {\bf 187}, 1767 (1969).


    
        \bibitem{hartmann_riedel} 
 B.~Hartmann and J.~Riedel,
  Phys.\ Rev.\ D {\bf 86} 104008 (2012). 
  
  
\bibitem{Herdeiro:2014goa} 
  C.~A.~R.~Herdeiro and E.~Radu,
  Phys.\ Rev.\ Lett.\  {\bf 112}, 221101 (2014).
  
  
\bibitem{Herdeiro:2015tia} 
C.~A.~R.~Herdeiro, E.~Radu and H.~Runarsson,
  Phys.\ Rev.\ D {\bf 92}, 084059 (2015).

 
\bibitem{Astefanesei:2003qy} 
  D.~Astefanesei and E.~Radu,
  Nucl.\ Phys.\ B {\bf 665}, 594 (2003)
 
  
\bibitem{Brihaye:2009yr} 
  Y.~Brihaye, T.~Caebergs, B.~Hartmann and M.~Minkov,
  Phys.\ Rev.\ D {\bf 80}, 064014 (2009).
  

  
\bibitem{Liebling:2012fv} 
  S.~L.~Liebling and C.~Palenzuela,
  Living Rev.\ Rel.\  {\bf 15}, 6 (2012).
    
\bibitem{Schunck:2003kk} 
  F.~E.~Schunck and E.~W.~Mielke,
  Class.\ Quant.\ Grav.\  {\bf 20}, R301 (2003).

 \bibitem{baseline} S. Doeleman, E. Agol, D. Backer, et al., {\it Astro2010: The Astronomy and Astrophysics Decadal Survey}, p. 68.

\bibitem{Lu:2014zja} 
  R.S. Lu, A.E. Broderick, F. Baron, J.D. Monnier, V.L. Fish, S.S. Doeleman, and V. Pankratius, Astrophys.\ J.\  {\bf 788}, 120 (2014).

\bibitem{Padilla:2012dx} 
  A.~Padilla and V.~Sivanesan,
  JHEP {\bf 1304}, 032 (2013).

    
    
\bibitem{Ohashi:2015fma} 
  S.~Ohashi, N.~Tanahashi, T.~Kobayashi and M.~Yamaguchi,
  JHEP {\bf 1507}, 008 (2015).
    
\bibitem{Kobayashi:2013ina} 
  T.~Kobayashi, N.~Tanahashi and M.~Yamaguchi,
  Phys.\ Rev.\ D {\bf 88}, no. 8, 083504 (2013).
  
\bibitem{Padilla:2010ir} 
  A.~Padilla, P.~M.~Saffin and S.~Y.~Zhou,
  Phys.\ Rev.\ D {\bf 83}, 045009 (2011).
  
\bibitem{Padilla:2010de} 
  A.~Padilla, P.~M.~Saffin and S.~Y.~Zhou,
  JHEP {\bf 1012}, 031 (2010).
  
  
\bibitem{Osilva} 
  M.~Horbatsch, H.~O.~Silva, D.~Gerosa, P.~Pani, E.~Berti, L.~Gualtieri and U.~Sperhake,
  Class.\ Quant.\ Grav.\  {\bf 32}, no. 20, 204001 (2015).
  
\bibitem{Cognola} 
 G.~Cognola, R.~Myrzakulov, L.~Sebastiani, S.~Vagnozzi and S.~Zerbini,
  arXiv:1601.00102 [gr-qc].

\bibitem{Saridakis:2016ahq} 
 E. N. Saridakis and M. Tsoukalas, Phys. Rev. D {\bf 93}, 124032 (2016).
  
\bibitem{Saridakis:2016mjd} 
 E. N. Saridakis and M. Tsoukalas, J. Cosmol. Astropart. Phys. 04 (2016) 017.
  
   
\bibitem{Derrick} G.~H.~Derrick, J.\ Math.\ Phys.\  {\bf 5}, 1252 (1964).
  
  \bibitem{colsys}
U. Ascher, J. Christiansen and R. D. Russell, Math. Comput. {\bf 33} 659
(1979);  ACM Trans. Math. Softw. {\bf 7} 209 (1981).
  

\bibitem{Pugliese:2013gsa} 
  D.~Pugliese, H.~Quevedo, J.~A.~Rueda H. and R.~Ruffini,
  Phys.\ Rev.\ D {\bf 88}, 024053 (2013).
\bibitem{Hartmann:2013tca}
  B.~Hartmann, J.~Riedel and R.~Suciu,
  Phys.\ Lett.\ B {\bf 726} 906 (2013). 
  
\bibitem{Brihaye:2015veu}
  Y.~Brihaye, A.~Cisterna, B.~Hartmann and G.~Luchini,
  Phys.\ Rev.\ D {\bf 92}, 124061 (2015).

   \bibitem{Alcubierre:2010ea} 
  M.~Alcubierre, J.~C.~Degollado, D.~Nunez, M.~Ruiz and M.~Salgado,
  Phys.\ Rev.\ D {\bf 81}, 124018 (2010).
 
\bibitem{Ruiz:2012jt} 
  M.~Ruiz, J.~C.~Degollado, M.~Alcubierre, D.~Nunez and M.~Salgado,
  Phys.\ Rev.\ D {\bf 86}, 104044 (2012).
 
 
\bibitem{Kleihaus:2015iea} 
  B.~Kleihaus, J.~Kunz and S.~Yazadjiev,
  Phys.\ Lett.\ B {\bf 744}, 406 (2015).



\end{thebibliography}

\end{document}